\documentclass[12pt]{article}

\usepackage[body={17cm, 23cm, centered}]{geometry}
\usepackage{epsfig,verbatim,cite}
\usepackage{lmodern}
\usepackage{braket}

\usepackage[usenames,dvipsnames,svgnames,table]{xcolor}
\usepackage{pgfplots}
\pgfplotsset{compat=1.18}
\usepgfplotslibrary{groupplots}

\usepackage[english]{babel}
\usepackage{titlesec}
\titleformat{\section}{\normalfont\Large\bfseries}{\thesection}{1em}{}

\usepackage{mathtext,marvosym,textcomp}
\usepackage{mathtools}
\usepackage{slashed}
\usepackage{tikz}
\usepackage{epsfig,amsmath,amsfonts,amssymb,arydshln}
\usepackage[makeroom]{cancel}
\usepackage{multirow}
\usepackage{tabularx}
\usepackage{graphicx}
\usepackage{float}

\usepackage[debug,pageanchor=false]{hyperref}
\hypersetup{colorlinks=true,linktocpage,breaklinks,
	urlcolor=blue,
	linkcolor=blue,
	citecolor=blue
}

\numberwithin{equation}{section}

\usepackage[pdftex]{pict2e}

\usepackage{tcolorbox}
\usetikzlibrary{positioning,fit,calc,arrows.meta}

\tikzset{
  modern/.style={
    rectangle,
    rounded corners=8pt,
    minimum width=3.5cm,
    minimum height=1.2cm,
    text centered,
    align=center,
    font=\sffamily\bfseries,
    fill=#1!15,
    draw=#1!80!black
  },
  arrowstyle/.style={
    -{Stealth[length=3mm,width=2mm]},
    thick,
    draw=black!85,
    rounded corners=8pt
  }
}

\begin{document}

\begin{titlepage}

\vspace{-3cm}

\begin{center}
\hfill ITEP/TH-32/26\\
\hfill IITP/TH-28/26\\
\hfill MIPT/TH-26/26\\
\end{center}

\vfill

\begin{center}
   \baselineskip=16pt
   {\large \bf Limits of the inverse scattering problem}
   \vskip 2cm
     Matvei Fedin${}^i{}^\pi{}^e$\footnote{\tt fedin.mm@iitp.ru, fedin.mm@phystech.edu },
     Kirill Gubarev${}^+{}^i{}^\pi$\footnote{\tt kirill.gubarev@phystech.edu }
     and Andrey Morozov$^i{}^\pi{}^e$\footnote{\tt morozov.andrey.a@iitp.ru}
       \vskip .6cm
             \begin{small}
                          {\it
                          $^i$Institute for Information Transmission Problems, 127051, Moscow, Russia\\
                          $^\pi$Moscow Institute of Physics and Technology, 
                          Laboratory of High Energy Physics, \\
                          9, Institutskii pereulok, 141702, Dolgoprudny, Russia\\
                          $^+$Institute of Theoretical and Mathematical Physics, Moscow State University, 119991, Russia\\
                          $^e$ITMO University, Kronverksky Prospect 49, Saint Petersburg 197101, Russia
                          } \\ 
\end{small}
\end{center}

\vfill 
\begin{center} 
\textbf{Abstract}
\end{center} 
\begin{quote}
The main goal of tomography is the reconstruction of density function out of its line integrals (integral measurements of this density along x-ray lines). Such construction is possible and known as inverse x-ray/Radon transformations. We are interested in generalization of this problem to the case of particles. For this we need to limit the speed of these particles, otherwise the problem is reduced to the previous one. The question we discuss in this paper is how low can this speed be depending on the parameters of the studied potential. We study this problem using simple theoretical examples and Machine Learning pipeline to restore the potential.
\end{quote}

\vfill
\setcounter{footnote}{0}
\end{titlepage}

\tableofcontents

\newpage

\setcounter{page}{2}

\section{Introduction}\label{sec:intro}

The reverse scattering problem known in practical fields as tomography is a long standing problem with lots of practical applications. Basically the idea is that we have some unknown medium which we cannot measure directly, but we can scatter waves or particles on it. Then we want to learn its shape and/or other properties based only on how it scatters incoming particles. This subject of course has lots of practical applications such as x-ray tomography in medicine and other fields \cite{ColtonKress2015,Faridani_Mathematical}.

The most basic case of reverse scattering problem is x-ray tomography. In this case we shoot the medium with x-ray waves and measure how effectively they can penetrate. It is know that in this case the problem can be solved by the Radon transformation if the dimension of the problem is at least two. We are interested in a more wide problem of reverse scattering.

We shoot the medium with some particles with known speed, from the known point at the border, and then measure when, where and with what speed they arrive. In this formulation the problem is actually not that interesting, since if we take the speed to be infinitely large we just get the x-ray problem again \cite{Jollivet_2006}. Therefore we want to limit speeds to some value. It is obvious that for zero speed and even for low speeds the problem is unsolvable since the particles cannot reach some parts of the studied medium. The question then is how low can this speed be for the potential to be still restorable.

It is known that for spherically symmetric repulsive potentials in classical mechanics, the differential scattering cross section at a single energy determines the potential outside the radius of closest approach, though higher energies are needed to probe closer to the center \cite{PhysRev.102.557}. For wave equations, similar reconstruction is possible via formal series expansions using reflection coefficients, and more generally the scattering operator can determine the potential if the diagonal representation is prescribed \cite{PhysRev.102.559}.

In the small-angle scattering regime (same as infinite energy limit), Novikov showed that scattering data at high energies uniquely determine the X-ray transforms of the force and potential, leading to full reconstruction of the force field for dimensions $d \geq 2$; for compactly supported potentials, this also yields a uniqueness theorem at fixed energy \cite{novikov1999small}. These results were later extended to the relativistic case by Jollivet, who demonstrated that the velocity-valued component of the scattering operator at high energies uniquely determines the X-ray transform, while the configuration-valued component does not \cite{Jollivet_2006}. Further work by Jollivet on inverse scattering and boundary value problems for relativistic and nonrelativistic Newton equations in static electromagnetic fields, building on the Gerver-Nadirashvili approach, established uniqueness theorems at sufficiently large fixed energy \cite{jollivet2007inverseproblemselectromagneticfield}.

In acoustic scattering, the classical result of Schiffer states that the far field pattern for all incident and observation directions at a fixed wave number uniquely determines the sound-soft obstacle. Rainer and collaborators showed that far fewer measurements --- specifically, far field data at isolated points --- can suffice for reconstruction, and provided constructive numerical methods \cite{Rainer}. In a different but related direction, Pestov and others proved that on a two-dimensional compact simple Riemannian manifold, knowledge of geodesic lengths between boundary points uniquely determines the metric up to natural obstruction; this was later generalized to higher dimensions for generic metrics, where one can check whether infinitesimal metric changes are recoverable from distance changes \cite{pestov2003dimensionalcompactsimpleriemannian, stefanov2004boundaryrigiditystabilitygeneric}. Muhometov's foundational work established uniqueness and stability for reconstructing a two-dimensional Riemannian metric in a domain from boundary distance data, a key result in geometric tomography \cite{muhometov1977}.

More recently, Machine Learning methods have been applied to inverse scattering problems. Adler and Oktem proposed the Learned Primal-Dual algorithm, which unrolls a proximal optimization method into a deep neural network trained end-to-end on raw data; it outperforms Filtered Back-Projection (FBP), Total Variation (TV), and learned post-processing in low-dose Computed Tomography (CT), achieving substantial Peak Signal-to-Noise Ratio (PSNR) and Structural Similarity Index Measure (SSIM) gains with only ten forward-back-projection steps \cite{Adler_2018}. Tsang and colleagues addressed the poor performance of Machine Learning (ML) methods in highly nonlinear scattering regimes by incorporating a differentiable forward solver as explicit physics knowledge, progressively refining reconstructions with increasing wave frequencies --- yielding high-quality results at reduced computational and sampling costs \cite{tsang2025modelguidedneuralnetworkmethod}.

Other works focus on potential reconstruction from phase shifts: Khachi and collaborators used the Variable Phase Approach with both Variational Monte Carlo (VMC) and multilayer perceptron (MLP) neural networks to construct local inverse potentials for deuteron states, obtaining nearly identical parameters while reducing the optimization to a one-dimensional problem \cite{khachi2024estimatinginversescatteringpotentials}. Neural networks have also been shown to solve inverse quantum problems such as finding a potential from its spectrum or from a prescribed many-electron density --- tasks traditionally considered much harder than the direct problem \cite{Lantz2021}. Finally, Jeong and others applied Physics-Informed Machine Learning (PIML), including Neural Ordinary Differential Equations (Neural ODEs) and Physics-Informed Neural Networks (PINNs), to inverse problems in holography and classical mechanics, reconstructing bulk spacetimes and effective potentials from boundary data; they also explored Kolmogorov-Arnold Networks (KANs) as a more efficient alternative, providing a systematic framework for using neural networks in inverse problems across physics and engineering \cite{jeong2025adsdeeplearningeasyiineural}.

We use Machine Learning algorithms to reconstruct the fields inside the studied medium, and we try to find out for which set of parameters the problem is solvable, and how low can the speed be.

The theoretical examples and numerical experiments suggest that reconstruction quality is governed by the ratio between the initial kinetic energy and a characteristic potential variation between the boundary and the interior of the domain. When these scales become comparable, the scattering map becomes strongly nonlinear and the reconstruction error begins to increase. 

From studied theoretical and computational examples our claim is that the minimal kinetic energy required to restore the potential is the modulo of the difference between potential energies inside and on the border:
\begin{equation}
E=\max(|U_{\text{in}}-U_{\text{out}}|).
\label{ans}
\end{equation}

The paper is structured as follows. In Section~\ref{sec:xray} we briefly recall the classical X-ray and Radon transforms. In Section~\ref{sec:newtonian-problem} we formulate the inverse scattering problem for Newtonian particles and develop a systematic expansion of the scattering data in inverse powers of the velocity. Two exactly solvable examples --- the harmonic potential and a constant force --- are studied analytically in Sections~\ref{sec:hp} and~\ref{sec:constf}, and a conjectural non-linear relation between the finite- and infinite-velocity ray integrals is discussed in Section~\ref{sec:generalJ}. Section~\ref{sec:ML} presents the machine-learning approach, including the architecture of the Force-Field Prediction Model (FFPM), the training procedure, and the numerical characterization of the applicability limit. We conclude and outline open questions in Section~\ref{sec:concl}. Entire learning pipeline were implemented in PyTorch with CUDA. The corresponding code and datasets are available at~\cite{Fedin2026}.

\subsection{Software and hardware}\label{subsec:software}

All learnable reconstruction models and training procedures were implemented in PyTorch. Paired training data and the evaluation-time physical-consistency tests were computed using a purpose-built CUDA solver for the Cauchy problem on the unit disk. Radon and filtered-backprojection reconstructions were evaluated on the GPU using an NVIDIA cuFFT implementation of the Ram--Lak filter and validated against the ASTRA Toolbox and scikit-image reference implementations \cite{astra,skimage}. All experiments were performed on a single machine; its software and hardware configurations are summarized in Tables~\ref{tab:software} and~\ref{tab:hardware}.

\begin{table}[htbp]
\centering
\footnotesize
\begin{minipage}[t]{0.52\textwidth}
  \centering
  \begin{tabularx}{\linewidth}{@{}l >{\raggedright\arraybackslash}X@{}}
  \hline
  Component & Version \\
  \hline
  OS & Debian 12 (Bookworm)\\
  Python & 3.11 \\
  PyTorch / torchvision & 2.4.1 / 0.19.1 (CUDA 12.1) \\
  CUDA runtime / compiler & 12.1 / 11.8 (nvcc) \\
    
  \hline
  \end{tabularx}
  \caption{Software stack.}
  \label{tab:software}
\end{minipage}\hfill
\begin{minipage}[t]{0.46\textwidth}
  \centering
  \begin{tabularx}{\linewidth}{@{}l >{\raggedright\arraybackslash}X@{}}
  \hline
  Component & Specification \\
  \hline
  CPU & AMD EPYC 7F52 (16 cores) \\
  RAM & 64\,GB \\
  GPU & $4 \times$ NVIDIA Tesla V100-SXM2 \\
  VRAM &128\,GB\\
  \hline
  \end{tabularx}
  \caption{Hardware stack.}
  \label{tab:hardware}
\end{minipage}
\end{table}

\section{X-ray and Radon transformation}\label{sec:xray}

The most well-known case of the inverse scattering problem is when we shoot light (X-ray) through the studied domain. The intensity of the light is then reduced based on the density of the medium. And it appears that through the measurement of intensity reduction along different trajectories one can find the density function of the medium.

Decrease of the intensity is described by an integral of the density along the trajectories, which are considered to be straight lines in this case:
\begin{equation}
Pf=\int\limits_L f \left(\mathbf{x}_{in}+t\boldsymbol{\tau}(\theta)\right) dt,
\label{xray}
\end{equation}
where $\mathbf{x}_{in}$ is an initial point of the trajectory and $\boldsymbol{\tau}(\theta)$ is a vector along the trajectory. This transformation is called an X-ray transformation. There is also a Radon transformation, which is analogous to the X-ray in two-dimensional case.
\begin{equation}
Rf(\rho,\theta)=\int f(\mathbf{x}) \delta(\mathbf{x} \cdot \mathbf{n}(\theta) -\rho) dxdy,
\label{Radon}
\end{equation}
where $\rho$ is an impact parameter, i.e. distance from the center of the area to the chord, $\mathbf{n}(\theta) \cdot \boldsymbol{\tau}(\theta) = 0$.

This transformation can be inversed in a Fourier-like way. First we need to build a Fourier transform of the Radon transformed function:
\begin{equation}
S(r,\theta)=\int Rf(\rho,\theta)e^{-2i\pi r\rho}d\rho.
\end{equation}
Then the function $f$ can be restored
\begin{equation}
f(x,y)=\int d\theta\int|r|S(r,\theta)e^{2i\pi r\rho}dr|_{\rho=x \cos\theta + y\sin\theta}
\end{equation}
This allows us to find the density of the medium using the X-ray image, and this approach is widely used in many practical applications.

\section{Newtonian mechanics and the force field}\label{sec:newtonian-problem}

In this paper we want to discuss similar, but different problem. We shoot Newtonian particles through the medium and want to understand what is the potential inside the studied area. Since we discuss the Newtonian mechanics there is a simple equation of motion:
\begin{equation}
m\ddot{\mathbf{x}}=\mathbf{F}(\mathbf{x}).
\end{equation}
Without loss of generality we can consider a particle of mass equal to $m=1$, and let us study the problem on a circle $B^2$ of radius 1. The story is different from the X-ray problem, since more of the variables change. In the X-ray case direction and speed remained the same through the whole transit of the ray through the medium. However, for particles all of these can change. Therefore we start from the initial conditions:
\begin{equation}
\dot{\mathbf{x}}=\mathbf{v},\ \ \ \mathbf{x}_0\in \partial B^2.
\end{equation}
Since the initial point lies on the boundary circle, it is parameterized by the polar angle \(\phi\). 

After transition of a particle we will get new properties and this gives us the whole set of measurable properties from which we can try to restore the potential inside the studied area:

\begin{enumerate}
\item initial boundary coordinate \(\phi\);
\item initial velocity \(\mathbf v\);
\item traversal time \(T(\phi,\mathbf v)\);
\item exit coordinate \(\Phi(\phi,\mathbf v)\);
\item exit velocity \(\mathbf V(\phi,\mathbf v)\).
\end{enumerate}

We use the force field $\mathbf F(\mathbf x)=-\nabla U(\mathbf x)$, rather than the scalar potential $U$, as the primary reconstruction variable. This eliminates the physically irrelevant ambiguity
\begin{equation}
U(\mathbf x)\longmapsto U(\mathbf x)+C,
\end{equation}
which does not affect either the force field or the particle trajectories.

The boundary launch point is parameterized by the polar angle \(\phi\):
\begin{equation}
\mathbf{x}_{in}(\phi)=(\cos\phi,\sin\phi).
\end{equation}
The angle \(\gamma\) is measured from the positively oriented tangent
\begin{equation}
\mathbf l(\phi)=(-\sin\phi,\cos\phi),
\end{equation}
whereas \(\alpha^{\mathrm{abs}}\) denotes the absolute direction of the initial velocity in the laboratory frame:
\begin{equation}
\alpha^{\mathrm{abs}}
=
\phi+\frac{\pi}{2}+\gamma.
\end{equation}

\begin{figure}[ht!]
\centering

\begin{tikzpicture}[
  >=Latex,
  vector/.style={
    ->,
    very thick
  },
  angle arc/.style={
    thick
  },
  guide/.style={
    gray,
    dashed,
    thin
  },
  every node/.style={
    font=\large
  }
]

  \def\R{3}
  \def\betaAngle{25}
  \def\gammaAngle{35}

  \pgfmathsetmacro{\alphaAngle}{
    \betaAngle + 90 + \gammaAngle
  }

  \coordinate (O) at (0,0);

  \coordinate (P) at (
    {\R*cos(\betaAngle)},
    {\R*sin(\betaAngle)}
  );

  \coordinate (V) at (
    {\R*cos(\betaAngle)+2.2*cos(\alphaAngle)},
    {\R*sin(\betaAngle)+2.2*sin(\alphaAngle)}
  );

  \draw[thick] (O) circle (\R);

  \fill (O) circle (1.5pt)
    node[below left] {$O$};

  \draw[guide]
    (O) -- (1.4,0);

  \draw[vector]
    (O) -- (P)
    node[midway,above left] {$\mathbf{x}_{in}(\phi)$};

  \fill (P) circle (1.7pt);

  \draw[guide]
    (P) -- ++(1.5,0);

  \draw[guide]
    (P) --
    ++(
      {1.45*cos(\betaAngle+90)},
      {1.45*sin(\betaAngle+90)}
    )
    node[pos=1.1,right] {$\mathbf{l}(\phi)$};

  \draw[vector,blue!70!black]
    (P) -- (V)
    node[pos=0.92,above left] {$\mathbf{v}$};

  \begin{scope}[shift={(O)}]

    \draw[angle arc]
      (0.72,0)
      arc[
        start angle=0,
        end angle=\betaAngle,
        radius=0.72
      ];

    \node at (
      {1.08*cos(\betaAngle/2)},
      {1.08*sin(\betaAngle/2)}
    ) {$\phi$};

  \end{scope}

  \begin{scope}[shift={(P)}]

    \draw[angle arc]
      (1.12,0)
      arc[
        start angle=0,
        end angle=\alphaAngle,
        radius=1.12
      ];

    \node at (
      {1.4*cos(\alphaAngle/4)},
      {1.4*sin(3*\alphaAngle/4)}
    ) {$\alpha^{\mathrm{abs}}$};

    \draw[angle arc]
      (
        {0.62*cos(\betaAngle+90)},
        {0.62*sin(\betaAngle+90)}
      )
      arc[
        start angle={\betaAngle+90},
        end angle=\alphaAngle,
        radius=0.62
      ];

    \node at (
      {0.95*cos(\betaAngle+90+\gammaAngle/2)},
      {0.95*sin(\betaAngle+90+\gammaAngle/2)}
    ) {$\gamma$};

  \end{scope}

\end{tikzpicture}

\caption{
Angular parametrization of the particle launch geometry. The angle \(\phi\) specifies the boundary launch point, \(\gamma\) is measured from the positively oriented tangent, and \(\alpha^{\mathrm{abs}}\) denotes the absolute laboratory-frame direction of the initial velocity.
}

\label{fig:scattering-geometry}
\end{figure}
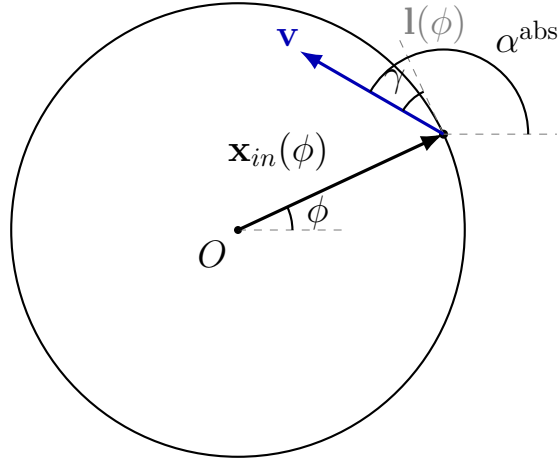

This angular convention is summarized in Fig.~\ref{fig:scattering-geometry}.

Now let us discuss how it can work in several examples. Let us start form the infinite-speed limit and see how Radon transformation can be used to restore the forces.

\subsection{Infinite-speed limit}\label{subsec:infspeed}

When speed is infinite the trajectory remains a straight line -- chord $L$, connecting $\phi$ and $\Phi$. We will also denote the length of this chord by $L$. The momentum of a particle will change, depending on a forces along the trajectory:
\begin{equation}
m \Delta \mathbf{v} = m (\mathbf{V}-\mathbf{v}) = \int\limits_0^T \mathbf{F}\left(\mathbf{x}(t)\right)dt=\frac{1}{v}\int\limits_0^L \mathbf{F}(\mathbf{x})dx + O(v^{-3}).
\end{equation}
We can define a vector ray integral along the chord:
\begin{equation}
\mathbf{J}_{\infty}(L)=\int\limits_0^L \mathbf{F}(\mathbf{x})dx=\lim\limits_{v\rightarrow\infty} \left(mv\Delta \mathbf{v}(L)\right).
\end{equation}
This formula actually is similar to the (\ref{Radon}), but it has two components, since it is defined for a vector. Therefore we can use the inverse Radon transformation to restore $\mathbf{F}$, just we need to use it kind of separately for two vector components:
\begin{equation}
\mathbf{F}(\mathbf{x})=\mathcal{R}^{-1}\left[\mathbf{J}_{\infty}\right]=\frac{1}{2\pi^2}\text{p.v.}\int\limits_0^{\pi}d\theta \int\limits_{-1}^1\cfrac{1}{\mathbf{x}\cdot\mathbf{n}(\theta)-\rho}\cfrac{\partial \mathbf{J}_{\infty}(L)}{\partial \rho}d\rho ,
\end{equation}
where chord $L$ is parametrized by again impact parameter $\rho$ and normal $\mathbf{n}(\theta)=(\cos\theta,\sin\theta)$, $\tau$ is the unit tangent of a chord $\boldsymbol{\tau}=(-\sin{\theta},\cos{\theta})$.

The existence of such a formula means that if we can shoot particles with infinite (or close to infinite) speeds then we can restore the forces inside the studied domain. However, if we make the speed lower this approach stops working.

\subsection{First order corrections}\label{subsec:finite-speed-expansion}

The question is how to move from the infinite speeds to finite speeds. If speed is finite the trajectory of the particle may no longer be the straight line. However we can try to calculate what are the corrections to the chord and to the measured momentum change due to the speed being finite. As we will show further the answer for the first corrections looks like
\begin{equation}
\mathbf{J}_v(L)=\mathbf{J}_{\infty}(L)+\frac{\delta U}{mv^2}\mathbf{Q}[\mathbf{F}](L)+\left(\frac{\delta U}{mv^2}\right)^2\mathbf{P}[\mathbf{F}](L)+O\left(\left(\frac{\delta U}{m v^2}\right)^3\right)
\end{equation}
where $\mathbf{Q}[\mathbf{F}]$ is a quadratic and $\mathbf{P}[\mathbf{F}]$ is a cubic functionals of $\mathbf{F}$ and its derivatives, integrated strictly along the unperturbed straight chord $L$ and $\delta U = \text{const}$ is characteristic value for the potential energy spread.

Let us now derive them via fixed boundary parameterization. Instead of allowing the spatial endpoints of the integration path to deform under finite velocity variations, we implement an alternative approach where the starting point $\mathbf{x}_{\text{in}}$ and the final point $\mathbf{x}_{\text{out}}$ on the boundary circle are kept strictly fixed. Let $L$ be the total length of the unperturbed straight chord connecting these two fixed points.

We introduce a normalized, dimensionless time variable $\chi = t / T$, where $T$ is the total time of flight for the true trajectory inside the disk. Consequently, $\chi \in [0, 1]$ is bounded by invariant limits. Newton's equation $m\ddot{\mathbf{x}} = \mathbf{F}(\mathbf{x})$ then looks like:
\begin{equation}\label{mainequation}
    m\frac{d^2\mathbf{x}}{d\chi^2} = T^2 \mathbf{F}(\mathbf{x})
\end{equation}

Both the time of flight $T$ and the spatial trajectory $\mathbf{x}(\chi)$ are expanded in perturbative power series with respect to $\epsilon = \delta U/(mv^2)$:
\begin{equation}
\begin{array}{lcl}
    T & = & \epsilon^{\frac12} T_0 + \epsilon^{\frac32} T_1 + \epsilon^{\frac52} T_2 + \ldots
\\ \\
    \mathbf{x}(\chi) & = & \mathbf{x}_0(\chi) + \epsilon \mathbf{y}_1(\chi) + \epsilon^2 \mathbf{y}_2(\chi) + \ldots
\end{array}
\end{equation}

The form of these expansions follows from a systematic iterative procedure. The unperturbed zero-order motion corresponds to a uniform traversal along the straight chord $L$:
\begin{equation}
    \mathbf{x}_0(\chi) = \mathbf{x}_{\text{in}} + \boldsymbol{\tau} L \chi, \qquad T_0 = L \sqrt{\frac{m}{\delta U}}.
\end{equation}
Since the physical entry and exit points are fixed, the perturbative coordinate deviations $\mathbf{y}_n(\chi)$ must satisfy homogeneous zero-boundary conditions at both ends of the scale:
\begin{equation}
    \mathbf{y}_n(0) = \mathbf{y}_n(1) = \mathbf{0}, \qquad \forall n \ge 1.
\end{equation}

Inserting the expansions into the equation of motion and the initial-speed condition $|\dot{\mathbf{x}}(0)| = v$, and collecting terms of the same order in $\epsilon$, generates a hierarchy of linear boundary-value problems.

\begin{itemize}
  \item At order $\epsilon$ one obtains the equation for $\mathbf{y}_1$ with a source term $\propto \mathbf{F}_0$, after that one finds the first correction $T_1$ fixed by the $\epsilon$ part of the speed condition. Hence $T_1$ appears at order $\epsilon^{3/2}$ in $T$, while $\mathbf{y}_1$ appears at order $\epsilon$ in $\mathbf{x}$.
  \item At order $\epsilon^2$ one solves for $\mathbf{y}_2$, whose source depends on $\mathbf{y}_1$ and $T_1$, and the condition on the speed determines $T_2$ (which enters $T$ at order $\epsilon^{5/2}$).
  \item This pattern repeats: at each step $n$ the deflection $\mathbf{y}_n$ is found from a pinned-end problem with a source built from the previously known $\mathbf{y}_1,\dots,\mathbf{y}_{n-1}$ and $T_1,\dots,T_{n-1}$, while $T_n$ is extracted from the $|\dot{\mathbf{x}}(0)|^2 = v^2$ condition at the corresponding order. The scaling guarantees that all $T_n$ enter as odd half-integer powers of $\epsilon$, and all $\mathbf{y}_n$ as integer powers.
\end{itemize}

Substituting the expansions into the equation of motion and collecting powers of
\(\epsilon\), we obtain the governing equations. Note that \(T^2 = \epsilon L^2 \frac{m}{\delta U} + 2\epsilon^2 L T_1 \sqrt{\frac{m}{\delta U}} + \epsilon^3 T_1^2 + 2\epsilon^3 L T_2 \sqrt{\frac{m}{\delta U}} + O(\epsilon^4)\).

\underline{Order \(\epsilon\):}
\begin{equation}
    m \mathbf{y}_1''(\chi) = L^2 \frac{m}{\delta U} \mathbf{F}_0(\chi), \qquad \mathbf{F}_0(\chi) = \mathbf{F}(\mathbf{x}_0(\chi)).
\end{equation}
Integrating with the fixed-endpoint conditions \(\mathbf{y}_1(0)=\mathbf{y}_1(1)=\mathbf{0}\) gives the Green's function solution
\begin{equation}
    \mathbf{y}_1(\chi) = \frac{L^2}{\delta U} \int_0^\chi (\chi - \eta) \mathbf{F}_0(\eta) \, d\eta - \frac{L^2}{\delta U} \chi \int_0^1 (1 - \eta) \mathbf{F}_0(\eta) \, d\eta.
\end{equation}

\underline{Order \(\epsilon^2 \):}
\begin{equation}
    m \mathbf{y}_2''(\chi) = L^2 \frac{m}{\delta U} (\mathbf{y}_1(\chi)\cdot\nabla)\mathbf{F}_0(\chi) + 2 L \sqrt{\frac{m}{\delta U}} T_1 \mathbf{F}_0(\chi),
    \qquad \mathbf{y}_2(0)=\mathbf{y}_2(1)=\mathbf{0}.
\end{equation}
Which can be solved similarly to the above equation:
\begin{multline}
    \mathbf{y}_2(\chi) = \frac{L^2}{\delta U} \int_0^\chi (\chi - \eta) \Bigl[ (\mathbf{y}_1(\eta)\,\cdot\,\nabla)\mathbf{F}_0(\eta)
    + 2\frac{T_1}{L}\mathbf{F}_0(\eta) \Bigr] \, d\eta \\
    - \frac{L^2}{\delta U} \chi \int_0^1 (1 - \eta) \Bigl[ (\mathbf{y}_1(\eta)\,\cdot\,\nabla)\mathbf{F}_0(\eta)
    + 2\frac{T_1}{L}\mathbf{F}_0(\eta) \Bigr] \, d\eta.
\end{multline}

We determine \(T_1\) and \(T_2\) from the condition that the magnitude of the initial velocity is exactly \(v\). The initial velocity is \(\mathbf{v}_{\text{in}} = T^{-1}\mathbf{x}'(0)\). Writing \(a = T_1/T_0\) and \(b = T_2/T_0\), the initial velocity is
\begin{equation}
  \mathbf{v}_{\text{in}} = v\Bigl[ \boldsymbol{\tau}
    + \frac{\epsilon}{L}\mathbf{y}_1'(0)
    + \frac{\epsilon^2}{L}\mathbf{y}_2'(0)
    - \epsilon a\boldsymbol{\tau}
    - \epsilon^2 a\frac{\mathbf{y}_1'(0)}{L}
    + \epsilon^2(a^2-b)\boldsymbol{\tau} + O(\epsilon^3)\Bigr].
\end{equation}
Imposing \(|\mathbf{v}_{\text{in}}|^2 = v^2\) order by order gives
\begin{equation}
a = \frac{1}{L}\boldsymbol{\tau}\!\cdot\!\mathbf{y}_1'(0),
\end{equation}
\begin{equation}
T_1 = T_0 a = \frac{T_0}{L}\,\boldsymbol{\tau}\!\cdot\!\mathbf{y}_1'(0)
    = -\frac{T_0 L}{\delta U}\int_0^1 (1-\eta)\,F_\parallel(\eta)\,d\eta
    = -\frac{T_0}{L\,\delta U}\int_0^L (L-u)\,F_\parallel(u)\,du.
\end{equation}
Here \(F_\parallel = \mathbf{F}_0\!\cdot\!\boldsymbol{\tau}\). The second-order condition yields
\begin{equation}
  b = \frac{1}{L}\boldsymbol{\tau}\!\cdot\!\mathbf{y}_2'(0) + \frac{1}{2}|\mathbf{V}_1|^2,
  \qquad
  \mathbf{V}_1 = \frac{1}{L}\mathbf{y}_1'(0) - a\boldsymbol{\tau}, \quad \boldsymbol{\tau}\,\cdot\, \mathbf{V}_1 = 0.
\end{equation}
Hence
\begin{equation}
  T_2 = T_0 b = \frac{T_0}{L}\,\boldsymbol{\tau}\!\cdot\!\mathbf{y}_2'(0) + \frac{T_0}{2L^2}\bigl(\mathbf{y}_{1,\perp}'(0)\bigr)^2,
\end{equation}
where \(\mathbf{y}_{1,\perp}'(0) = \mathbf{y}_1'(0)\!\cdot\!\mathbf{n}\) is the normal component of the first-order endpoint derivative.

The physical momentum change is obtained directly from Newton's second law:
\begin{equation}
m \Delta \mathbf{v} = \int_0^T \mathbf{F}(\mathbf{x}(t))\,dt = T\int_0^1 \mathbf{F}(\mathbf{x}(\tau))\,d\tau.
\end{equation}
The force integral becomes
\begin{equation}
\int_0^1 \mathbf{F}(\mathbf{x}(\chi))\,d\chi
= \int_0^1 \mathbf{F}_0\,d\chi
  + \epsilon\!\int_0^1 (\mathbf{y}_1\!\cdot\!\nabla)\mathbf{F}_0\,d\chi
  + \epsilon^2\!\int_0^1 \Bigl[ (\mathbf{y}_2\!\cdot\!\nabla)\mathbf{F}_0
      + \frac12(\mathbf{y}_1\!\cdot\!\nabla)^2\mathbf{F}_0 \Bigr]d\chi
  + O(\epsilon^3).
\end{equation}
Now we multiply the above expression by \(vT\).  Using \(v = \sqrt{\delta U/(m\epsilon)}\) and the expansion of \(T\),
\begin{equation}
vT = \sqrt{\frac{\delta U}{m\epsilon}}\Bigl(\epsilon^{1/2}T_0 + \epsilon^{3/2}T_1 + \epsilon^{5/2}T_2 + \cdots\Bigr)
    = L + \epsilon\sqrt{\frac{\delta U}{m}}\,T_1 + \epsilon^2\sqrt{\frac{\delta U}{m}}\,T_2 + O(\epsilon^3).
\end{equation}
Carrying out the product and switching to the arc length \(s = L\tau\) (\(ds = L\,d\tau\)), we obtain
\begin{equation}\label{eq:Jv_eps2}
  \mathbf{J}_v(L) = \mathbf{J}_\infty(L)
  + \frac{\delta U}{m v^2}\,\mathbf{Q}[\mathbf{F}](L)
  + \left(\frac{\delta U}{m v^2}\right)^2\mathbf{P}[\mathbf{F}](L)
  + O\left(\left(\frac{\delta U}{m v^2}\right)^3\right),
\end{equation}
where the first- and second-order functionals are
\begin{equation}
\begin{aligned}
\mathbf{Q}[\mathbf{F}](L) &=
   \int_0^L (\mathbf{y}_1(s)\!\cdot\!\nabla)\mathbf{F}_0(s)\,ds
   + \frac{\sqrt{\delta U/m}\;T_1}{L}\,\mathbf{J}_\infty(L),\\[4pt]
\mathbf{P}[\mathbf{F}](L) &=
   \int_0^L (\mathbf{y}_2(s)\!\cdot\!\nabla)\mathbf{F}_0(s)\,ds
   + \frac12\int_0^L (\mathbf{y}_1(s)\!\cdot\!\nabla)^2\mathbf{F}_0(s)\,ds\\
   &\qquad + \frac{\sqrt{\delta U/m}\;T_1}{L}\int_0^L (\mathbf{y}_1(s)\!\cdot\!\nabla)\mathbf{F}_0(s)\,ds
   + \frac{\sqrt{\delta U/m}\;T_2}{L}\,\mathbf{J}_\infty(L).
\end{aligned}
\end{equation}
Inserting \(\mathbf{y}_1\), \(\mathbf{y}_2\) and \(T_1\), \(T_2\) into \(\mathbf{Q}[\mathbf{F}]\), \(\mathbf{P}[\mathbf{F}]\) yields an explicit, purely force-dependent functional that contains up to second derivatives of the field.  For a general potential this expression is rather lengthy, but it simplifies dramatically for the harmonic oscillator, where \(\mathbf{Q}[\mathbf{F}] = 0\) and \(\mathbf{P}[\mathbf{F}]\) evaluates to a non-zero constant, see Section \ref{sec:hp}. 

Now we have a systematic expansion of the measured vector ray integral in powers of \(\delta U/(m v^2)\):
\begin{equation}
\mathbf{J}_v(L) = \mathbf{J}_\infty(L) + \frac{\delta U}{m v^2}\,\mathbf{Q}[\mathbf{F}](L)
                + \left(\frac{\delta U}{m v^2}\right)^2\mathbf{P}[\mathbf{F}](L)
                + O\left(\left(\frac{\delta U}{m v^2}\right)^3\right),
\end{equation}
where \(\mathbf{J}_\infty(L)=\mathcal{R}[\mathbf{F}](L)\) is the classical Radon transform of the true force field, and the functionals \(\mathbf{Q}\) and \(\mathbf{P}\) are given above.  Applying the inverse Radon operator \(\mathcal{R}^{-1}\) which is linear to the measured data yields an \emph{effective} (naively reconstructed) field
\begin{equation}
\mathbf{F}_v(\mathbf{x}) \equiv \mathcal{R}^{-1}[\mathbf{J}_v](\mathbf{x})
= \mathbf{F}(\mathbf{x}) + \frac{\delta U}{m v^2}\,\mathbf{G}_1[\mathbf{F}](\mathbf{x})
                     + \left(\frac{\delta U}{m v^2}\right)^2\mathbf{G}_2[\mathbf{F}](\mathbf{x})
                     + O(v^{-6}),
\end{equation}
with
\begin{equation}
\mathbf{G}_1[\mathbf{F}] = \mathcal{R}^{-1}\bigl[\mathbf{Q}[\mathbf{F}]\bigr], \qquad
\mathbf{G}_2[\mathbf{F}] = \mathcal{R}^{-1}\bigl[\mathbf{P}[\mathbf{F}]\bigr].
\end{equation}
Explicitly, using the inversion formula,
\begin{multline}
    \mathbf{F}_v(\mathbf{x}) = \mathbf{F}(\mathbf{x})
      + \frac{1}{2\pi^2} \frac{\delta U}{m v^2} \,
        \mathrm{p.v.}\!\int_{0}^{\pi}\!\!d\theta \int_{-1}^{1}\,
        \frac{\partial_\rho \mathbf{Q}[\mathbf{F}](\theta,\rho)}{\mathbf{x}\!\cdot\!\mathbf{n}(\theta)-\rho}\, d\rho \\[4pt]
      + \frac{1}{2\pi^2} \left(\frac{\delta U}{m v^2}\right)^2\,
        \mathrm{p.v.}\!\int_{0}^{\pi}\!\!d\theta \int_{-1}^{1}\,
        \frac{\partial_\rho \mathbf{P}[\mathbf{F}](\theta,\rho)}{\mathbf{x}\!\cdot\!\mathbf{n}(\theta)-\rho}\, d\rho
      + O(v^{-6}).
\end{multline}
If the speed is infinite, \(\mathbf{J}_v=\mathbf{J}_\infty\) and \(\mathbf{F}_v=\mathbf{F}\), then the classical Radon inversion formula gives the exact force field. For large but finite \(v\) the same formula still provides a first approximation to \(\mathbf{F}\); the error is \(O\left(\left(\frac{\delta U}{m v^2}\right)^2\right)\).

Retaining only the \(O\!\left(\frac{\delta U}{m v^2}\right)\) term we have
\begin{equation}
\mathbf{F}_v = \mathbf{F} + \frac{\delta U}{m v^2}\,\mathbf{G}_1[\mathbf{F}] + O\!\left(\Bigl(\frac{\delta U}{m v^2}\Bigr)^2\right).
\end{equation}
This relation can be used in two ways:
\begin{itemize}
  \item \textbf{Single-speed iteration:} Starting from \(\mathbf{F}^{(0)}=\mathbf{F}_v\),
        define
        \begin{equation}
        \mathbf{F}^{(k+1)} = \mathbf{F}_v - \frac{\delta U}{m v^2}\,\mathbf{G}_1\bigl[\mathbf{F}^{(k)}\bigr].
        \end{equation}
        The correction is of order \(\epsilon^2\) after the first iteration, so the method converges rapidly provided \(v\) exceeds a critical threshold \(v_c\).  The critical speed is determined by the condition that the dimensionless parameter \(\epsilon = \frac{\delta U}{m v^2}\) is smaller than a constant \(\epsilon_c\) that depends on the strength of the force field \(\mathbf{F}\) and the size of the domain (for the harmonic oscillator with \(\delta U=1,\ m=1\) one finds \(\epsilon_c=1/2\), i.e.\  \(v_c=\sqrt{2}\))\footnote{This critical speed is equal to $\sqrt{2\delta U m}=\sqrt{2}$, where $\delta U$ and $m$ are taken to be $1$ in some units of measurement. It has \textbf{no} relation to natural units of measurement where speed of light is equal to $1$. Units of measurement can be easily added if rescaling of the answers is needed.}.
  \item \textbf{Two-speed extrapolation:} If measurements at two speeds \(v_1\neq v_2\) are available, one can eliminate \(\mathbf{G}_1\) algebraically and obtain \(\mathbf{F}\) directly. Writing \(\mathbf{F}_{v_i} = \mathbf{F} + \frac{\delta U}{m v_i^2}\mathbf{G}_1 + \cdots\), subtraction gives
        \begin{equation}
        \mathbf{G}_1 = \frac{\mathbf{F}_{v_1} - \mathbf{F}_{v_2}}{\frac{\delta U}{m}\bigl(v_1^{-2}-v_2^{-2}\bigr)},
        \qquad
        \mathbf{F} = \mathbf{F}_{v_1} - \frac{\delta U}{m v_1^2}\mathbf{G}_1.
        \end{equation}
        With more than two speeds one performs a linear regression in \(1/v^2\) and extrapolates to \(1/v^2\to0\) (renormalisation group flow).  This procedure does not require prior knowledge of \(\mathbf{G}_1\) and automatically works whenever the expansion is valid.
\end{itemize}

If the \(O\!\left(\bigl(\frac{\delta U}{m v^2}\bigr)^2\right)\) term is also retained, the full equation becomes
\begin{equation}
\mathbf{F}_v = \mathbf{F} + \frac{\delta U}{m v^2}\mathbf{G}_1[\mathbf{F}]
                     + \left(\frac{\delta U}{m v^2}\right)^2\mathbf{G}_2[\mathbf{F}]
                     + O\!\left(\Bigl(\frac{\delta U}{m v^2}\Bigr)^3\right).
\end{equation}
The iteration scheme is extended to
\begin{equation}
\mathbf{F}^{(k+1)} = \mathbf{F}_v
   - \frac{\delta U}{m v^2}\mathbf{G}_1\bigl[\mathbf{F}^{(k)}\bigr]
   - \left(\frac{\delta U}{m v^2}\right)^2\mathbf{G}_2\bigl[\mathbf{F}^{(k)}\bigr].
\end{equation}
Alternatively, with three or more speeds one can perform a quadratic regression in \(1/v^2\) and extrapolate to zero.  The procedure can be repeated order-by-order, yielding a systematic reconstruction of the force field from finite-speed boundary data.  The critical speed for the extended scheme is essentially the same as for the first-order one, because the small parameter is still \(\epsilon = \frac{\delta U}{m v^2}\).

\section{Example: harmonic potential}\label{sec:hp}
\subsection{Scattering geometry}\label{subsec:hpgeometry}

Consider the harmonic potential $U(r)=r^2$, $\mathbf{F}=-2\mathbf{x}$ on the unit disk. A particle of mass $m=1$ is launched from the boundary point $(x_0,y_0)=(1,0)$ ($\phi=0$) with speed $v$, the velocity vector making an angle $\alpha$ with the inward radial direction ($-\pi/2\le\alpha<\pi/2$) in general but due to the symmetry of a potential with respect of $\alpha \to -\alpha$ we will consider ($0\le\alpha<\pi/2$) \footnote{Here and throughout this section, $\alpha$ denotes the local angle with the inward radial direction and must not be confused with the absolute laboratory-frame angle $\alpha^{\mathrm{abs}}_{ij}$ used in the numerical discretization.}.  The motion terminates when the particle returns to the boundary $r=1$; let the final polar angle be $\Phi$. The equations of motion decouple into two harmonic oscillators with frequency $\omega=\sqrt2$:
\begin{equation}
\ddot x+2x=0,\qquad \ddot y+2y=0 .
\end{equation}
With the initial conditions $x(0)=1$, $y(0)=0$, $\dot x(0)=-v\cos\alpha$,
$\dot y(0)=v\sin\alpha$, the solutions are
\begin{equation}
x(t)=\cos(\sqrt2 t)-\frac{v\cos\alpha}{\sqrt2}\sin(\sqrt2 t),\qquad
y(t)=\frac{v\sin\alpha}{\sqrt2}\sin(\sqrt2 t).
\end{equation}
Eliminating $t$ and converting to polar coordinates $(\varphi,r)$ yields the elliptic orbit
\begin{equation}
\frac1{r^2(\varphi)}=A+B\cos 2\varphi+C\sin 2\varphi,
\end{equation}
with coefficients
\begin{equation}
A=\frac1{2\sin^2\alpha}\!\left(1+\frac2{v^2}\right),\quad
B=\frac12-\frac1{2\sin^2\alpha}\!\left(\cos^2\alpha+\frac2{v^2}\right),\quad
C=\cot\alpha .
\end{equation}
The ellipse is then rotated by an angle $\theta_0$ given by
\begin{equation}
\tan2\theta_0=\frac{C}{B}= \frac{\sin2\alpha}{-\cos2\alpha-2/v^2}.
\end{equation}
Then the boundary $r=1$ is reached at the initial point ($\phi=0$) and at the final point $\Phi=2\theta_0$, whence the exact scattering relation
\begin{equation}\label{eq:tanphi}
\tan\Phi = \frac{\sin2\alpha}{-\cos2\alpha-2/v^2}.
\end{equation}
In the infinite-speed limit $v\to\infty$ this reduces to $\Phi=\pi-2\alpha$, corresponding to the straight chord of the Radon transform.

The map $\alpha\mapsto\Phi$ exhibits a sharp change at the critical velocity $v_c=\sqrt2$, intimately related to the convergence of the $1/v$ series. There are three different modes for this system:

\begin{itemize}
\item\textbf{Low velocities ($v<\sqrt2$).}  The denominator in (\ref{eq:tanphi}) never vanishes; the map $\Phi$ has a minimum on $(0,\pi/2)$, and the image is a restricted arc $[\Phi_{\min},\pi]$, $\Phi_{\text{min}}=\arccos\!\left(-\frac{\sqrt{4 - v^{4}}}{2}\right)$.  Every accessible boundary point (except the minimum) is hit by exactly two distinct injection angles.

\item\textbf{High velocities ($v>\sqrt2$).}  The denominator can cross zero, the map becomes monotonic and the image covers the full circle $[0,2\pi)$.  Thus the scattering is a bijection: every boundary point is reached exactly once.

\item\textbf{Critical value ($v=\sqrt2$).}  At this velocity a tangentially launched particle stays on the rim, $\Phi\to 0$, and the transition from two-to-one to one-to-one coverage occurs.
\end{itemize}

Therefore we can see that the potential can be restored when
\begin{equation}
    E=\frac{mv^2}{2}>1=U(1)-U(0).
\end{equation}

\subsection{Vector ray integral and its exact form}\label{subsec:hpJ}

We work in the coordinate system attached to the chord $L$. Let the chord have normal $\mathbf{n}=(\cos\theta,\sin\theta)$, tangent $\boldsymbol{\tau}=(-\sin\theta,\cos\theta)$, and impact parameter $\rho$ from the origin. Its length is $L=2\sqrt{1-\rho^2}$.  The entry and exit points are
\begin{equation}
\mathbf{x}_{\rm in}=\rho\mathbf{n}-\frac{L}{2}\boldsymbol{\tau},\qquad
\mathbf{x}_{\rm out}=\rho\mathbf{n}+\frac{L}{2}\boldsymbol{\tau}.
\end{equation}

\textbf{Infinite-speed limit.}  For $v\to\infty$ the trajectory is the straight chord $L$, and the vector ray integral is simply the line integral of the force $\mathbf{F}=-2\mathbf{x}$:
\begin{equation}
\mathbf{J}_\infty(L)=\int_L \mathbf{F}\,ds
   =\int_{-L/2}^{L/2} (-2)\bigl(\rho\mathbf{n}+s\boldsymbol{\tau}\bigr)\,ds
   = -4\rho\sqrt{1-\rho^2}\,\mathbf{n}
   \equiv -2s\,\mathbf{n},
\end{equation}
where we introduced the convenient dimensionless parameter $s\equiv 2\rho\sqrt{1-\rho^2}$.  For the particular injection studied earlier, $\rho=\sin\alpha$ and $s=\sin2\alpha$, recovering $\mathbf{J}_\infty=-2\sin2\alpha\,\mathbf{n}$.

\textbf{Finite speed.}  For a given $v$ the particle follows a true trajectory satisfying $\ddot{\mathbf{x}}=-2\mathbf{x}$. The general solution with frequency $\omega=\sqrt2$ is
\begin{equation}
\mathbf{x}(t)=\mathbf{A}\cos\omega t+\mathbf{B}\sin\omega t .
\end{equation}
Imposing $\mathbf{x}(0)=\mathbf{x}_{\rm in}$ and $\mathbf{x}(T)=\mathbf{x}_{\rm out}$ fixes
\begin{equation}
\mathbf{A}= \mathbf{x}_{\rm in},\qquad
\mathbf{B}= \frac{\mathbf{x}_{\rm out}-\mathbf{x}_{\rm in}\cos\omega T}{\sin\omega T}.
\end{equation}
Using the explicit coordinates of the endpoints,
\begin{equation}
\mathbf{x}_{\rm out}-\mathbf{x}_{\rm in}\cos\omega T
= (1-\cos\omega T)\rho\,\mathbf{n} + \frac{L}{2}(1+\cos\omega T)\boldsymbol{\tau}.
\end{equation}
Hence the initial velocity is $\mathbf{v}_{\rm in}=\dot{\mathbf{x}}(0)=\omega\mathbf{B}$, and its modulus squared becomes
\begin{equation}
v^2 = \omega^2\,\frac{(1-\cos\omega T)^2\rho^2 + \frac{L^2}{4}(1+\cos\omega T)^2}{\sin^2\omega T}.
\end{equation}
Using $L^2=4(1-\rho^2)$ and and introducing notation $t \equiv \tan(\omega T/2)$, the expression for $v^2$ simplifies to
\begin{equation}\label{eq:v2_t}
v^2 = 2\Bigl( \rho^2 t^2 + \frac{1-\rho^2}{t^2} \Bigr).
\end{equation}
Thus $t^2$ satisfies the quadratic equation
\begin{equation}
2\rho^2 t^4 - v^2 t^2 + 2(1-\rho^2) = 0 .
\end{equation}
Its discriminant is $\Delta = v^4 - 16\rho^2(1-\rho^2) = v^4 - 4s^2$, where $s = 2\rho\sqrt{1-\rho^2}$.  The physical root that tends to zero as $v\to\infty$ is
\begin{equation}
t^2 = \frac{v^2 - \sqrt{v^4 - 4s^2}}{4\rho^2}.
\end{equation}

Introduce the angle $\beta$ via $\sin\beta = 2s/v^2$ (so $\beta = \arcsin(2s/v^2)$). Then $\sqrt{v^4-4s^2} = v^2\cos\beta$ and
\begin{equation}
t^2 = \frac{v^2 - v^2\cos\beta}{4\rho^2}
    = \frac{v^2(1-\cos\beta)}{4\rho^2}
    = \frac{v^2\sin^2(\beta/2)}{2\rho^2},
\end{equation}
which yields
\begin{equation}
t = \frac{v\,\sin(\beta/2)}{\sqrt2\,\rho}.
\end{equation}

The change in momentum $\Delta\mathbf{v} = \mathbf{v}_{\rm out} - \mathbf{v}_{\rm in}$ can be evaluated directly from the harmonic solution. The normal component is the only that survives:
\begin{equation}
\Delta\mathbf{v} = -2\omega \rho\,t\,\mathbf{n}
                = -2\sqrt2\,\rho\,t\,\mathbf{n}.
\end{equation}
Multiplying by $v$ and substituting the above expression for $t$ gives the exact finite-speed ray integral
\begin{equation}\label{eq:Jv_exact}
\mathbf{J}_v(L) = v\Delta\mathbf{v}
   = -2v\sqrt2\,\rho\,\frac{v\sin(\beta/2)}{\sqrt2\,\rho}\,\mathbf{n}
   = -2v^2\sin\!\Bigl(\frac12\arcsin\frac{2s}{v^2}\Bigr)\,\mathbf{n}
   = -\frac{2}{\varepsilon}\,
     \sin\!\Bigl(\frac12\arcsin(2\varepsilon s)\Bigr)\,\mathbf{n},
\end{equation}
with $\varepsilon=1/v^2$.  In the limit $\varepsilon\to0$ the sine behaves as $\varepsilon s$, and $\mathbf{J}_v\to -2s\,\mathbf{n}=\mathbf{J}_\infty$, perfectly matching the Radon-transform result.

\subsection{Reconstruction at finite speed: explicit verification}\label{subsec:hpRadon}

All odd powers of $1/v$ vanish in the series for $\mathbf{J}_v$; in particular the $O(v^{-2})$ correction $\mathbf{Q}[\mathbf{F}]$ is identically zero. The first non-trivial term is of order $1/v^4$ and equals $\frac18\mathbf{J}_\infty^3$. Hence
\begin{equation}
\mathbf{J}_v(L) = \mathbf{J}_\infty(L) + \frac{1}{8v^4}\mathbf{J}_\infty^3(L) + O(v^{-8}),
\qquad \mathbf{J}_\infty(L) = -2s\,\mathbf{n}.
\end{equation}

The series converges absolutely for $|x| = 2s/v^2 \le 1$, i.e.\ for $v \ge \sqrt{2s}$. The maximal value $s_{\max}=1$ gives the critical velocity $v_c=\sqrt2$, exactly the value found from the topological analysis.  For $v>v_c$ the series converges uniformly for all chords; for $v<v_c$ it diverges on a set of chords and the perturbative treatment breaks down.

We now apply the inverse vector Radon transform $\mathcal{R}^{-1}$ (which is linear) to the truncated series.  By definition $\mathcal{R}^{-1}[\mathbf{J}_\infty] = \mathbf{F}$, and for the harmonic oscillator $\mathbf{F}(\mathbf{x}) = -2\mathbf{x}$.  Using the exact form $\mathbf{J}_\infty(L) = -2s\,\mathbf{n}$ with $s = 2\rho\sqrt{1-\rho^2}$, the cubic term $\mathbf{J}_\infty^3$ is a purely normal vector field
\begin{equation}
\mathbf{J}_\infty^3(L) = -8\,s^3\,\mathbf{n} = -64\,\rho^3(1-\rho^2)^{3/2}\,\mathbf{n}.
\end{equation}
To compute its inverse Radon transform we first evaluate
\begin{equation}
\mathcal{R}^{-1}\bigl[\rho^3(1-\rho^2)^{3/2}\,\mathbf{n}\bigr](\mathbf{x})  = \frac{1}{2\pi^2 }\mathrm{p.v.}\!\int_{0}^{\pi}\!\!d\theta \int_{-1}^{1}\, \frac{\partial_\rho (\rho^3(1-\rho^2)^{3/2})}{\mathbf{x}\cdot\mathbf{n}(\theta)-\rho}\, \mathbf{n}(\theta) d\rho .
\end{equation}
Inserting this function into the inversion formula (2.1) and performing the angular integration (the principal-value integral reduces to a standard Hilbert transform) yields
\begin{equation}
\mathcal{R}^{-1}\bigl[\rho^3(1-\rho^2)^{3/2}\,\mathbf{n}\bigr](\mathbf{x}) = \frac{3}{16}(1-r^{2})(5r^{2}-1)\,\mathbf{x} = - \frac{3}{16}(1-6r^{2} + 5r^{4})\,\mathbf{x}.
\end{equation}
Consequently,
\begin{equation}
\mathcal{R}^{-1}\bigl[\mathbf{J}_\infty^3\bigr](\mathbf{x}) = -64\cdot\frac{3}{16}(-1+6r^{2} - 5r^{4})\,\mathbf{x} = 12(1-6r^{2} + 5r^{4})\,\mathbf{x}.
\end{equation}
The naively reconstructed field at finite speed is therefore
\begin{multline}
\mathbf{F}_v(\mathbf{x}) = \mathcal{R}^{-1}[\mathbf{J}_v](\mathbf{x})
   = \mathcal{R}^{-1}[\mathbf{J}_\infty](\mathbf{x}) + \frac{1}{8v^4}\mathcal{R}^{-1}\bigl[\mathbf{J}_\infty^3\bigr](\mathbf{x}) + O(v^{-8}) \\
   = -2\mathbf{x} + \frac{3}{2v^4}\,(1-6r^2+5r^4)\,\mathbf{x} + O(v^{-8}).
\end{multline}

We now compare this with the general perturbative relation:
\begin{equation}
\mathbf{F}_v(\mathbf{x}) = \mathbf{F}(\mathbf{x}) + \frac{1}{v^2}\mathbf{G}_1[\mathbf{F}](\mathbf{x})
                     + \frac{1}{v^4}\mathbf{G}_2[\mathbf{F}](\mathbf{x}) + O(v^{-6}).
\end{equation}
For the harmonic force $\mathbf{F} = -2\mathbf{x}$, direct computation yields $\mathbf{Q}[\mathbf{F}] = 0$, hence $\mathbf{G}_1[\mathbf{F}] = 0$, and $\mathbf{P}[\mathbf{F}] = \frac18\mathbf{J}_\infty^3$, so that
\begin{equation}
\mathbf{G}_2[\mathbf{F}] = \frac18\mathcal{R}^{-1}\bigl[\mathbf{J}_\infty^3\bigr]
   = \frac{3}{2}\,(1-6r^2+5r^4)\,\mathbf{x}.
\end{equation}
Substituting these values into the general equation reproduces exactly the expression for $\mathbf{F}_v$ obtained from the exact series. This explicit check confirms the internal consistency of the perturbative framework.

The true force field $\mathbf{F} = -2\mathbf{x}$ is now recovered from the measured $\mathbf{F}_v$ by solving the above relation.  Because $\mathbf{G}_1=0$, the equation simplifies to
\begin{equation}
\mathbf{F}_v = \mathbf{F} - \frac{3}{4v^4}\,(1-6r^2+5r^4)\,\mathbf{F} + O(v^{-8}),
\end{equation}
which can be inverted to give
\begin{equation}
\mathbf{F}(\mathbf{x}) = \frac{\mathbf{F}_v(\mathbf{x})}{1 - \dfrac{3}{4v^4}\,(1-6r^2+5r^4)}
   = -2\mathbf{x} + O(v^{-8}).
\end{equation}
Equivalently, a single iteration of the general reconstruction scheme,
\begin{equation}
\mathbf{F}^{(1)}(\mathbf{x}) = \mathbf{F}_v(\mathbf{x}) - \frac{1}{8v^4}
   \mathcal{R}^{-1}\!\bigl[(\mathcal{R}[\mathbf{F}_v])^3\bigr](\mathbf{x}) + \cdots,
\end{equation}
cancels the $v^{-4}$ correction up to $O(v^{-8})$.  Thus the true harmonic field is recovered with high precision, demonstrating the practicability of the finite-speed Radon inversion.

\section{Example: constant force}\label{sec:constf}

Consider a uniform force field $\mathbf{F}(\mathbf{x}) = \mathbf{g}$, $\mathbf{g}\in\mathbb R^2$, on the unit disk. The potential is linear: $U(\mathbf{x}) = -\mathbf{g}\!\cdot\!\mathbf{x}$. We follow the same fixed-boundary setup as in the previous section: a chord $L$ is given by its normal $\mathbf{n}(\theta)$ and impact parameter $\rho$, $L=2\sqrt{1-\rho^2}$; the entry and exit points are
\begin{equation}
\mathbf{x}_{\rm in}=\rho\mathbf{n}-\frac{L}{2}\boldsymbol{\tau},\qquad
\mathbf{x}_{\rm out}=\rho\mathbf{n}+\frac{L}{2}\boldsymbol{\tau}.
\end{equation}
The particle is launched with speed $v$ from $\mathbf{x}_{\rm in}$ and must hit $\mathbf{x}_{\rm out}$ in the true trajectory.

\subsection{Exact vector ray integral}\label{subsec:constfgeom}

The equation of motion $\ddot{\mathbf{x}} = \mathbf{g}$ is solved by
\begin{equation}
\mathbf{x}(t) = \mathbf{x}_{\rm in} + \mathbf{v}_{\rm in} t + \frac12 \mathbf{g} t^2 .
\end{equation}
Imposing $\mathbf{x}(T)=\mathbf{x}_{\rm out}$ gives
\begin{equation}
\mathbf{v}_{\rm in} = \frac{L}{T}\boldsymbol{\tau} - \frac12 \mathbf{g} T .
\end{equation}
The initial speed $v$ is fixed, hence
\begin{equation}\label{eq:const_v2}
v^2 = \frac{L^2}{T^2} - L\,g_\parallel + \frac{g^2}{4}\,T^2 ,
\end{equation}
where $g_\parallel = \mathbf{g}\!\cdot\!\boldsymbol{\tau}$, $g_\perp = \mathbf{g}\!\cdot\!\mathbf{n}$, $g = |\mathbf{g}|$. Multiplying by $T^2$ yields a quadratic for $T^2$:
\begin{equation}
\frac{g^2}{4} T^4 - (v^2 + L g_\parallel) T^2 + L^2 = 0 .
\end{equation}
The physically relevant root (which reduces to $T = L/v$ when $\mathbf{g}=\mathbf 0$) is
\begin{equation}
T^2 = \frac{2}{g^2}\Bigl( v^2 + L g_\parallel - \sqrt{(v^2+L g_\parallel)^2 - g^2 L^2} \,\Bigr) .
\end{equation}
The change of momentum is $\Delta\mathbf{v} = \mathbf{g} T$, so the measured ray integral is
\begin{equation}
\mathbf{J}_v(L) = v\,\Delta\mathbf{v} = vT\,\mathbf{g}.
\end{equation}
Using the above expression for $T$ one obtains the exact closed form
\begin{equation}\label{eq:Jv_const}
\mathbf{J}_v(L) = \frac{\sqrt{2}\,v^2}{g}\,
\sqrt{\, v^2 + L g_\parallel - \sqrt{(v^2+L g_\parallel)^2 - g^2 L^2} }\;
\frac{\mathbf{g}}{g} \; .
\end{equation}
In the infinite-speed limit $v\to\infty$ we have $T \to L/v$, hence
\begin{equation}
\mathbf{J}_\infty(L) = L\,\mathbf{g},
\end{equation}
which is precisely the Radon transform of the constant force $\mathbf{F}=\mathbf{g}$. This confirms that the classical inversion formula of Section~\ref{subsec:infspeed} recovers $\mathbf{g}$ from $\mathbf{J}_\infty$.

\subsection{Series expansion and verification of the perturbative formulas}\label{subsec:constfRadon}

Expanding (\ref{eq:Jv_const}) in powers of $1/v^2$ gives
\begin{equation}
\mathbf{J}_v(L) = L\mathbf{g} + \frac{L^2}{2v^2}\,g_\parallel\,\mathbf{g}
                + \frac{L^3}{8v^4}\bigl(3g_\parallel^2 - g_\perp^2\bigr)\,\mathbf{g}
                + O(v^{-6}) .
                \label{eq:Jconst}
\end{equation}
Thus the first-order correction is
\begin{equation}
\mathbf{Q}[\mathbf{F}](L) = \frac{L^2}{2}\,g_\parallel\,\mathbf{g}.
\end{equation}
This agrees with the general formula evaluated for $\mathbf{F}=\mathbf{g}$: a direct calculation yields $\mathbf{y}_1(s) = \frac12 s(L-s)\mathbf{g}$, $F_\parallel = g_\parallel$, $\mathbf{J}_\infty = L\mathbf{g}$, and $\int_0^L (\mathbf{y}_1\!\cdot\!\nabla)\mathbf{F}_0\,ds = 0$, leaving $-\frac{\mathbf{J}_\infty}{L}\int_0^L (L-u)g_\parallel\,du = \frac{L^2}{2}g_\parallel\mathbf{g}$. Similarly, the second-order term matches $\mathbf{P}[\mathbf{F}]$.

The series converges whenever the square root in (\ref{eq:Jv_const}) is analytic, i.e.\ when the discriminant is positive: $(v^2+L g_\parallel)^2 > g^2 L^2$. This requires
\begin{equation}
v^2 > L\,(g - g_\parallel) .
\end{equation}
The right-hand side depends on the chord; its maximum over all chords defines a critical velocity $v_c = \max_{\rho,\theta} \sqrt{L\,(g - g_\parallel)}=\sqrt{2Lg}$. For $v>v_c$ the expansion converges uniformly.

\section{Hypothetical general form for $\mathbf{J}_\infty$}\label{sec:generalJ}

Based on the exactly solvable examples we discussed, one may conjecture that for a wide class of central or constant forces the exact relation between $\mathbf{J}_\infty$ and $\mathbf{J}_v$ has the universal structure
\begin{equation}\label{eq:hypothetical}
\mathbf{J}_\infty(L) =
\frac{2\left(1 - \dfrac{|\mathbf{J}_v|^2}{4v^4}\right)}
     {\sqrt{4 - \dfrac{(\mathbf{J}_v\cdot\mathbf{n})^2}{v^4}} -
      \dfrac{\mathbf{J}_v\cdot\boldsymbol{\tau}}{v^2}}
\; \mathbf{J}_v(L)\;.
\end{equation}
For the harmonic oscillator $\mathbf{J}_v$ is purely normal, i.e.\ $\mathbf{J}_v\!\cdot\!\boldsymbol{\tau} = 0$, and (\ref{eq:hypothetical}) collapses to the simple multiplicative correction
\begin{equation}
\mathbf{J}_\infty(L) = \mathbf{J}_v(L)\,
\sqrt{\,1 - \frac{|\mathbf{J}_v|^2}{4v^4}\,}.
\end{equation}
Let us show how that works.
\begin{equation}
\mathbf{J}_v(L) = -\frac{2}{\varepsilon}\,\sin\!\Bigl(\frac12\arcsin(2\varepsilon s)\Bigr)\,\mathbf{n},
\qquad \varepsilon = \frac{1}{v^2},\quad s = 2\rho\sqrt{1-\rho^2},
\end{equation}
implies the local relation $\mathbf{J}_\infty = \mathbf{J}_v\sqrt{1-|\mathbf{J}_v|^2/(4v^4)}$. Denote $J_v = |\mathbf{J}_v| = \frac{2}{\varepsilon}\sin\bigl(\frac12\arcsin(2\varepsilon s)\bigr)$. Set $y = \frac12\arcsin(2\varepsilon s)$, so that $\sin y = \varepsilon J_v/2$. Then
\begin{equation}
2\varepsilon s = \sin(2y) = 2\sin y\cos y = 2\sin y\sqrt{1-\sin^2 y}
= \varepsilon J_v\sqrt{1 - \frac{\varepsilon^2 J_v^2}{4}} .
\end{equation}
Hence $s = \frac{1}{2}J_v\sqrt{1 - J_v^2/(4v^4)}$.
Since $\mathbf{J}_\infty = -2s\,\mathbf{n}$ and $\mathbf{J}_v = -J_v\,\mathbf{n}$, we obtain
\begin{equation}
\mathbf{J}_\infty = -J_v\sqrt{1 - \frac{J_v^2}{4v^4}}\;\mathbf{n}
= \mathbf{J}_v\sqrt{1 - \frac{|\mathbf{J}_v|^2}{4v^4}} .
\end{equation}
This is exactly the hypothetical formula (\ref{eq:hypothetical}) for the special case $\mathbf{J}_v\!\cdot\!\boldsymbol{\tau}=0$, valid for the harmonic oscillator.

Just as for the harmonic oscillator for the constant force case, one can eliminate the force $\mathbf{g}$ and the chord parameters from the exact solution to obtain a direct local relation between the measured $\mathbf{J}_v$ and the ideal $\mathbf{J}_\infty$. Using the identity $\mathbf{J}_\infty = L\mathbf{g}$ and the fact that $\mathbf{J}_v$ is parallel to $\mathbf{g}$, a straightforward algebraic inversion of (\ref{eq:Jv_const}) also leads to (\ref{eq:hypothetical}). It is an exact consequence of the equations of motion for a constant force and provides a non-linear correction that can be applied directly to the data without any knowledge of $\mathbf{g}$.

For a general potential formula (\ref{eq:hypothetical}) is not exact, but it may serve as a valuable first approximation when non-local effects are small.


\section{Machine learning approach}\label{sec:ML}

\subsection{Numerical forward model and data acquisition}\label{sec:ml-forward-model}

Throughout this section, \(\widehat{X}\) denotes the model prediction of a ground-truth quantity \(X\). The learning experiments use the Newtonian inverse problem defined in Section~\ref{sec:newtonian-problem}. The learned reconstructor predicts the conservative force field \(\widehat{\mathbf F}\) directly from the boundary measurements. When a scalar representation is required for evaluation or visualization, the corresponding potential \(\widehat{U}\) is recovered by least-squares Fourier integration and is fixed up to an additive constant.

For $i,j=0,\ldots,255$, the numerical launch geometry introduced in Section~\ref{sec:newtonian-problem} is discretized as
\begin{equation}
\begin{array}{ll}
\phi_i=\frac{2\pi i}{256},
&
\mathbf{x}_{in}(\phi_i)=(\cos\phi_i,\sin\phi_i),\\
\mathbf l(\phi_i)=(-\sin\phi_i,\cos\phi_i),
&
\gamma_j=\frac{\pi j}{256},\\
\alpha^{\mathrm{abs}}_{ij}
=\phi_i+\frac{\pi}{2}+\gamma_j,
&
\mathbf v_{ij}
=v_0\left(
\cos\alpha^{\mathrm{abs}}_{ij},
\sin\alpha^{\mathrm{abs}}_{ij}
\right).
\end{array}
\end{equation}

The initial speed is fixed at $v_0=1$, while the amplitude $A$ of the potential is varied. After nondimensionalization, the trajectory geometry depends on the ratio $A/v_0^2$; varying $A$ at fixed $v_0$ therefore probes the same relative dynamical regimes as varying $v_0$ for a fixed normalized potential shape.

Trajectories are integrated using the fourth-order Runge--Kutta method with time step $dt=10^{-4}$ and maximum integration time $t_{\max}=10$. The force field is represented on a $256\times256$ grid and evaluated using CUDA texture interpolation.

\subsection{Data}\label{sec:ml-data}

\subsubsection{Encoding the input and target field}\label{sec:ml-encoding}

For each initial condition, the simulator returns the final velocity \(\mathbf v^{\,\mathrm{out}}\) and the number of steps \(N_{\mathrm{in}}\) during which the particle was inside the region. These form the measuring tensor \(D\in[-1,1]^{3\times256\times256}\): 

\begin{equation}
\begin{array}{lcl}
D_1&=&\cfrac{\mathbf{V}_y}{\left|\mathbf V\right|+10^{-12}},\\ \\
D_2&=&2\cfrac{\left|\mathbf{V}\right|}{v_0}-1,\\ \\
D_3&=&2\cfrac{N_{\mathrm{in}}}{N_{\mathrm{steps}}}-1.
\end{array}
\label{eq:measurement-channels}
\end{equation}

Values outside the nominal range are clipped to the corresponding endpoint.

Thus, the first channel contains information about the direction of the final velocity, the second about the change in its magnitude, and the third about the particle's residence time within the region, i.e., the time between the launch and the recrossing of the boundary. The spatial coordinates of the exit point are not included directly in the ML input.

In the conservative formulation, the potential value at the boundary is related to channel \(D_2\) by the energy conservation law. Therefore, \(D_2\) has little information content, while \(D_3\) reflects the change in the residence time of the particle within the region. The physical interpretation of the channels, however, does not determine their actual contribution to the prediction; this issue is explored in Section~\ref{sec:ml-input-information}.

In addition to the measurement tensor \(D\), the reconstructor receives an approximate Radon reconstruction \(R\), which serves as a physically motivated prior. The model uses the nonlinear boundary-scattering measurements to correct the large-scale structure provided by this initial approximation. The three measurement channels and the three Radon reconstruction channels form the six-channel input, while the two-component target field is represented using three channels:

\begin{equation}
X=\operatorname{concat}(D,R)\in\mathbb{R}^{6\times256\times256}, \qquad Y=(0,F_y,F_x).
\label{eq:ffpm-input-target}
\end{equation}

The first target channel is reserved for compatibility with the RGB representation. The model predicts the conservative force field rather than the scalar potential, thereby avoiding the physically irrelevant additive constant of the latter.

\subsubsection{Distribution of synthetic fields}\label{sec:ml-synthetic-fields}

As will be seen below, the problem largely corresponds to the Data-Centric AI paradigm, so one of the main ways to build a successful model is to construct a highly diverse dataset. In our problem, we combined several smooth (to circumvent the potential gradient explosion problem) potential types. Each of them may have specific patterns that introduce bias into the model's predictions, but taken together, these biases are in opposite directions and suppress the effect of each.

For each scalar potential \(U\), the corresponding conservative force field is
\begin{equation}
\mathbf{F}(\mathbf{x})=-\boldsymbol{\nabla} U(\mathbf{x}).
\end{equation}

To generate potentials, we use Gaussian fields with RBF covariance, see Appendix~\ref{app:field-rbf}, Mat\'ern fields, \ref{app:field-matern}, truncated Fourier series~\ref{app:field-fourier}, mixtures of Gaussians~\ref{app:field-gmm}, and Perlin noise~\ref{app:field-perlin}, as well as sums of compact harmonic~\ref{app:field-compact} and anisotropic Gaussian wells~\ref{app:field-anisotropic}, making the dataset diverse enough to reduce the likelihood of OOD inference and allow for more generalizable conclusions. The potentials undergo a series of transformations and smoothing to ensure the stability of numerical experiments; a full definition of this pipeline is given in the appendix~\ref{app:field-processing}.

The main series of experiments considers $16$ potential amplitude values:

\begin{equation}
    A_k=\frac{1}{10}\cdot2^{k-5},
\qquad k=0,\ldots,15.
\label{eq:ffpm-amplitude-grid}
\end{equation}

For each amplitude, $2000$ samples are generated independently, meaning the complete set contains $32000$ ``measurement --- field'' pairs. The data is divided into training and validation portions at a ratio of $90/10$.

\subsection{Reconstructor architecture}\label{sec:ml-architecture}

The reconstructor maps the six-channel tensor \(X\) defined in \eqref{eq:ffpm-input-target}, consisting of boundary measurements and a Radon reconstruction, into a three-channel representation of the force field:
\begin{equation}
\widehat{\mathbf F}=G_\theta(X)
\in\mathbb{R}^{3\times256\times256}.
\end{equation}
Further, we will call this model the \emph{Force-Field Prediction Model} (FFPM). The potential is reconstructed, if necessary, from the predicted force field at the postprocessing stage; the corresponding procedure is given in Section~\ref{sec:ml-metrics}.

The FFPM is based on a Johnson-style generator with the architecture encoder $\rightarrow$ residual bottleneck $\rightarrow$ decoder \cite{johnson2016perceptual}. During training, the generator is augmented with two unconditional PatchGAN-style discriminators operating at different spatial scales \cite{isola2017pix2pix,wang2018pix2pixhd}.  The detailed structure is given in the appendix~\ref{app:reconstructor-architecture} and shown in Fig.~\ref{fig:ml-model-architecture}.

The input measurements are parameterized by the initial-condition coordinates of the initial conditions \((\phi,\gamma)\), while the reconstructed field is defined on a Cartesian grid \((x,y)\). Therefore, identical positions of the input and output tensors do not correspond to the same physical point. The long encoder--decoder skip connections characteristic of U-Net \cite{ronneberger2015unet} imply a more direct spatial correspondence between features of different levels. In the problem under consideration, such a correspondence is absent, so U-Net-like schemes were not included in the final comparison, and the transformation is performed through a pronounced bottleneck.

\begin{figure}[htbp]
\centering
\definecolor{racegreen}{rgb}{0.1725,0.6275,0.1725}
\definecolor{raceblue}{rgb}{0.1216,0.4667,0.7059}
\definecolor{raceorange}{rgb}{1.0000,0.4980,0.0549}
\definecolor{racepurple}{rgb}{0.5804,0.4039,0.7412}
\definecolor{racered}{rgb}{0.8392,0.1529,0.1569}
\definecolor{racegray}{rgb}{0.5020,0.5020,0.5020}
\begin{tikzpicture}
\begin{groupplot}[
  group style={group size=3 by 1,horizontal sep=5mm},
  width=6.25cm,
  height=6.65cm,
  grid=both,
  grid style={line width=.25pt,draw=gray!22},
  tick align=outside,
  tick pos=left,
  tick label style={font=\scriptsize},
  label style={font=\scriptsize},
  title style={font=\small},
  every axis plot/.append style={line width=.9pt,mark size=1.25pt},
  legend cell align=left,
  legend style={font=\tiny,draw=gray!50,fill=white,fill opacity=0.88,
    text opacity=1,rounded corners=1pt,at={(0.98,0.98)},anchor=north east}
]
\nextgroupplot[title={(a) Model race},xlabel={time (min)},
  ylabel={diagnostic $\operatorname{relMAE}(F)$}]
\addplot[color=racegreen,mark=*] coordinates {(0,0.335) (2.3,0.296) (4.4,0.283) (7.2,0.258) (10.1,0.258) (12.6,0.245) (16,0.214) (18.5,0.231) (20.8,0.214) (23.2,0.182) (25.4,0.196) (27.6,0.153) (30.1,0.15) (32.5,0.155) (34.8,0.159)};
\addlegendentry{GAN 9rb: .150}
\addplot[color=raceblue,mark=*] coordinates {(0,0.52) (3.3,0.357) (6.8,0.336) (9.4,0.313) (11.6,0.307) (14.8,0.304) (17.6,0.306) (20.1,0.268) (22.5,0.249) (24.2,0.248) (26.3,0.256) (28.8,0.236) (30.9,0.231) (33.3,0.233)};
\addlegendentry{DiT-S: .231}
\addplot[color=raceorange,mark=*] coordinates {(0,0.455) (3.3,0.423) (7,0.394) (9.7,0.361) (12,0.347) (14.5,0.329) (17.7,0.373) (20.1,0.335) (22.7,0.328) (24.8,0.337) (27.1,0.383) (29.4,0.338) (31.2,0.331) (33,0.32)};
\addlegendentry{DiT-S HB: .320}
\addplot[color=racepurple,mark=*] coordinates {(0,0.36) (8,0.331) (16.5,0.31) (23.2,0.287) (29.7,0.27)};
\addlegendentry{DiT-B: .270}

\nextgroupplot[title={(b) GAN variants},xlabel={time (min)}]
\addplot[color=racegreen,mark=*] coordinates {(0,0.317) (1.4,0.286) (2.5,0.286) (3.4,0.266) (4.4,0.276) (5.7,0.242) (6.9,0.242) (8.1,0.231) (9.2,0.253) (10.5,0.216) (11.6,0.231) (12.7,0.221) (14.1,0.195) (15.7,0.191) (16.9,0.177) (18,0.173) (19,0.182) (20,0.25) (21.2,0.159) (22.4,0.151) (23.6,0.159)};
\addlegendentry{9rb: .151}
\addplot[color=raceblue,mark=*] coordinates {(0,0.34) (1.2,0.305) (2.1,0.282) (3.2,0.268) (4.1,0.263) (5.4,0.25) (6.4,0.246) (7.7,0.242) (8.7,0.235) (9.6,0.215) (10.4,0.222) (11.3,0.225) (12.6,0.187) (13.8,0.191) (15,0.192) (16.5,0.19) (17.5,0.179) (18.7,0.164) (19.7,0.166) (20.5,0.162) (21.6,0.169) (22.7,0.147)};
\addlegendentry{9rb comp.: .147}
\addplot[color=raceorange,mark=*] coordinates {(0,0.375) (1.2,0.425) (2.1,0.277) (3,0.259) (3.9,0.251) (4.8,0.257) (6,0.242) (6.9,0.225) (8,0.226) (9,0.215) (10.2,0.219) (11,0.217) (12,0.2) (13.1,0.185) (14.3,0.213) (15.5,0.222) (16.6,0.175) (17.6,0.174) (18.5,0.175) (19.6,0.16) (20.8,0.147) (21.7,0.155) (22.7,0.162) (23.7,0.152)};
\addlegendentry{6rb: .147}
\addplot[color=racered,mark=*] coordinates {(0,0.302) (1.2,0.377) (2,0.295) (3,0.263) (3.8,0.25) (4.7,0.259) (5.7,0.248) (6.5,0.229) (7.6,0.253) (8.5,0.229) (9.3,0.213) (10.1,0.2) (10.9,0.198) (11.8,0.184) (12.7,0.18) (13.7,0.171) (14.7,0.163) (15.7,0.169) (16.8,0.157) (17.7,0.155) (18.5,0.154) (19.5,0.152) (20.5,0.141) (21.4,0.136) (22.3,0.131) (23.3,0.124)};
\addlegendentry{6rb comp.: .124}

\nextgroupplot[title={(c) Capacity},xlabel={amplitude $A$},xmode=log]
\addplot[color=racegray,mark=square*] coordinates {(0.05,0.129) (0.2,0.233) (0.8,0.43) (3.2,0.809)};
\addlegendentry{6rb / 150 ep.}
\addplot[color=racegreen,mark=*] coordinates {(0.05,0.131) (0.2,0.235) (0.8,0.419) (3.2,0.773)};
\addlegendentry{9rb / 250 ep.}
\end{groupplot}
\end{tikzpicture}
\caption{Selecting a working configuration: (a) time-to-quality comparison of residual
FFPM and DiT variants with the same wall-clock budget and validation protocol;
(b) preliminary comparison of FFPM depth and \texttt{torch.compile} mode;
(c) jointly varying training depth and duration. All
configurations were run once, so the results are used as an
engineering criterion for model selection and do not characterize variability
with respect to initialization.}
\label{fig:ml-model-selection}
\end{figure}
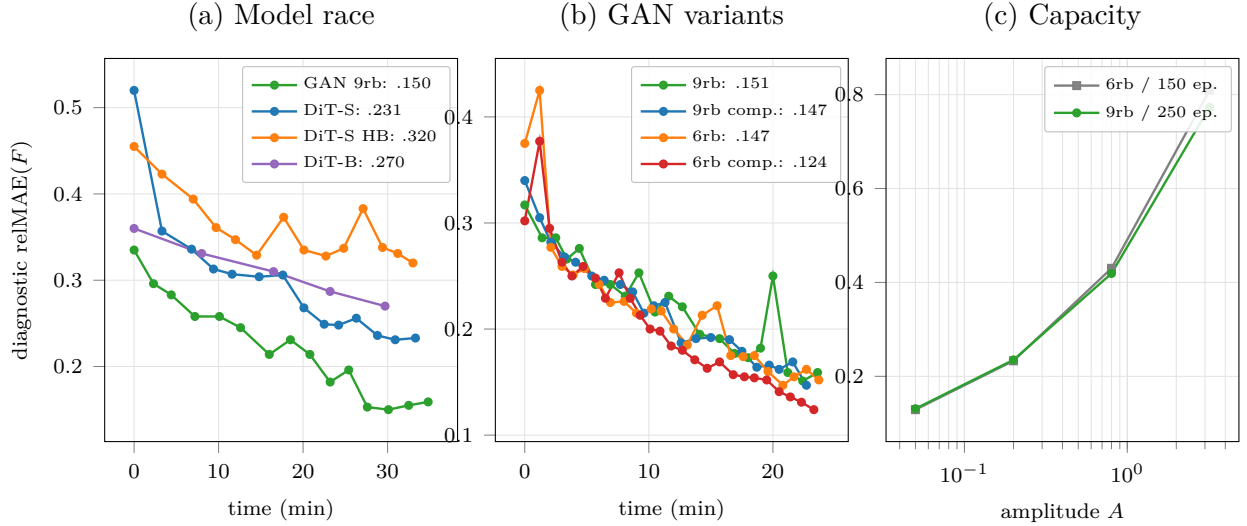

During the design phase, Johnson-style residual models were compared with diffusion transformers from the DiT family \cite{peebles2023dit}. Fig.~\ref{fig:ml-model-selection} compares their training speed and also examines the depth and capacity of the residual model.

In the experiments considered, Johnson-style FFPM achieved lower errors faster than the tested DiT configurations. Increasing the number of residual blocks from six to nine and the training duration did not result in a systematic improvement in performance. Therefore, the faster configuration with six residual blocks was chosen for the main experiments.

\subsection{Metrics}\label{sec:ml-metrics}

The primary metric for force recovery is the relative mean absolute error.
\begin{equation}
\operatorname{relMAE}(F)=
\frac{\left\langle|\widehat{\mathbf F}-\mathbf F|\right\rangle_{\Omega_{0.85}}}
{\left\langle|\mathbf F|\right\rangle_{\Omega_{0.85}}},
\qquad
\Omega_{0.85}={\mathbf x:|\mathbf x|\le0.85},
\end{equation}
where restricting the inner region of the disk reduces the influence of boundary and interpolation effects.

Additionally, the relative \(L_2\) error, Pearson correlation, and multiplicative gain are used:

\begin{equation}
\operatorname{relL2}(F)=
\frac{|\widehat{\mathbf F}-\mathbf F|_2}
{|\mathbf F|_2},
\qquad \rho(X,Y)=
\frac{
\left\langle
(X-\langle X\rangle)(Y-\langle Y\rangle)
\right\rangle
}{
\sqrt{
\left\langle(X-\langle X\rangle)^2\right\rangle
\left\langle(Y-\langle Y\rangle)^2\right\rangle
}
}, \qquad
\operatorname{gain}(F)=
\frac{\langle\widehat{\mathbf F},\mathbf F\rangle}
{\langle\mathbf F,\mathbf F\rangle}.
\end{equation}

To estimate the potential component, the field is projected onto gradient fields in Fourier space with the zero mode of the potential fixed. The relative error of the potential is defined as
\begin{equation}
\operatorname{relRMS}(U)=
\sqrt{
\frac{
\left\langle
\left[(\widehat U-U)-\langle\widehat U-U\rangle\right]^2
\right\rangle
}{
\left\langle
\left[U-\langle U\rangle\right]^2
\right\rangle
}
}.
\end{equation}

For the closed-loop metric, the number of successful repeated simulations is specified separately; unsuccessful runs are excluded from the average.

\subsection{Loss function and training}\label{sec:ml-training}

\begin{figure}[ht!]
\centering
\resizebox{0.94\linewidth}{!}{%
\begin{tikzpicture}[node distance=12mm and 15mm]

  \node (measurements) [
    modern=yellow,
    minimum width=4.2cm
  ] {
    Boundary measurements\\
    $D\in\mathbb{R}^{3\times256\times256}$
  };

  \node (reconstruction) [
    modern=yellow,
    right=of measurements,
    minimum width=4.2cm
  ] {
    Radon reconstruction\\
    $R\in\mathbb{R}^{3\times256\times256}$
  };

  \node (truth) [
    modern=yellow,
    right=of reconstruction
  ] {
    True field\\
    $\mathbf F$
  };

  \node (data) [
    draw=black!65,
    dashed,
    rounded corners=8pt,
    fit=(measurements)(truth)(reconstruction),
    inner sep=11pt,
    label={
      [font=\sffamily\bfseries]
      left:Training data
    }
  ] {};

  \coordinate (inputmid)
    at ($(measurements)!0.5!(reconstruction)$);

  \node (concat) [
    modern=teal,
    below=14mm of inputmid,
    minimum width=6.2cm
  ] {
    $X=\operatorname{concat}_{\mathrm{ch}}(D,R)\quad X
    \in\mathbb{R}^{6\times256\times256}$
  };

  \node (generator) [
    modern=green,
    below=of concat,
    minimum width=5.2cm
  ] {
    FFPM $G_\theta$
  };

  \node (prediction) [
    modern=teal,
    below=of generator,
    minimum width=5.2cm
  ] {
    $\widehat{\mathbf F}=G_\theta(X)$
  };

  \draw[arrowstyle]
    (measurements.south)
    --
    (measurements.south |- concat.north);

  \draw[arrowstyle]
    (reconstruction.south)
    --
    (reconstruction.south |- concat.north);

  \draw[arrowstyle] (concat) -- (generator);
  \draw[arrowstyle] (generator) -- (prediction);

  \node (nce) [
    modern=green,
    right=25mm of generator,
    yshift=9mm
  ] {
    $\mathcal L_{\mathrm{NCE}}$
  };

  \node (rec) [
    modern=green,
    right=of nce
  ] {
    $\mathcal L_{\mathrm{rec}}$
  };

  \node (gan) [
    modern=green,
    below=5mm of nce
  ] {
    $\mathcal L_{\mathrm{GAN}}$
  };

  \node (lossgroup) [
    draw=black!65,
    dashed,
    rounded corners=8pt,
    fit=(rec)(nce)(gan),
    inner sep=11pt,
    label={
      [font=\sffamily\bfseries]
      right:Loss terms
    }
  ] {};

  \draw[arrowstyle]
    (truth.south)
    -- ++(0,-5mm)
    -| (rec.north);

  \draw[arrowstyle]
    (reconstruction.south)
    -- ++(0,-5mm)
    -| (nce.north);

   \draw[arrowstyle]
    (prediction.east)
    --
    node[
      above,
      font=\sffamily\scriptsize
    ] {
    }
    ++(8mm,0)
    |- (lossgroup.west);

  \node (optimizer) [
    modern=green,
    below=of prediction,
    minimum width=5.4cm
  ] {
    linearly decaying learning rate
  };

\path (optimizer -| lossgroup) coordinate (posTR);

\node[modern=red,
      minimum width=8.3cm] (totalloss) at (posTR)
{
  $\mathcal L_G
  =
  20\mathcal L_{\mathrm{rec}}
  +0.2\mathcal L_{\mathrm{NCE}}
  +0.5\mathcal L_{\mathrm{GAN}}$
};

  \draw[arrowstyle] (lossgroup) -- (totalloss);

  \draw[arrowstyle] (totalloss) -- (optimizer);

  \draw[arrowstyle,dashed]
    (optimizer.west)
    --
    ([xshift=-8mm]generator.west |- optimizer.west)
    --
    ([xshift=-8mm]generator.west)
    --
    node[
      above,
      font=\sffamily\scriptsize
    ] {
      parameter update
    }
    (generator.west);

  \node (valpred) [
    modern=teal,
    below=of optimizer,
    minimum width=5.4cm
  ] {
    Validation $\widehat{\mathbf F}_{\mathrm{EMA}}$
  };

  \draw[arrowstyle] (optimizer) -- (valpred);

\end{tikzpicture}%
}

\caption{
FFPM training and evaluation pipeline.
}
\label{fig:ml-training-evaluation-pipeline}
\end{figure}
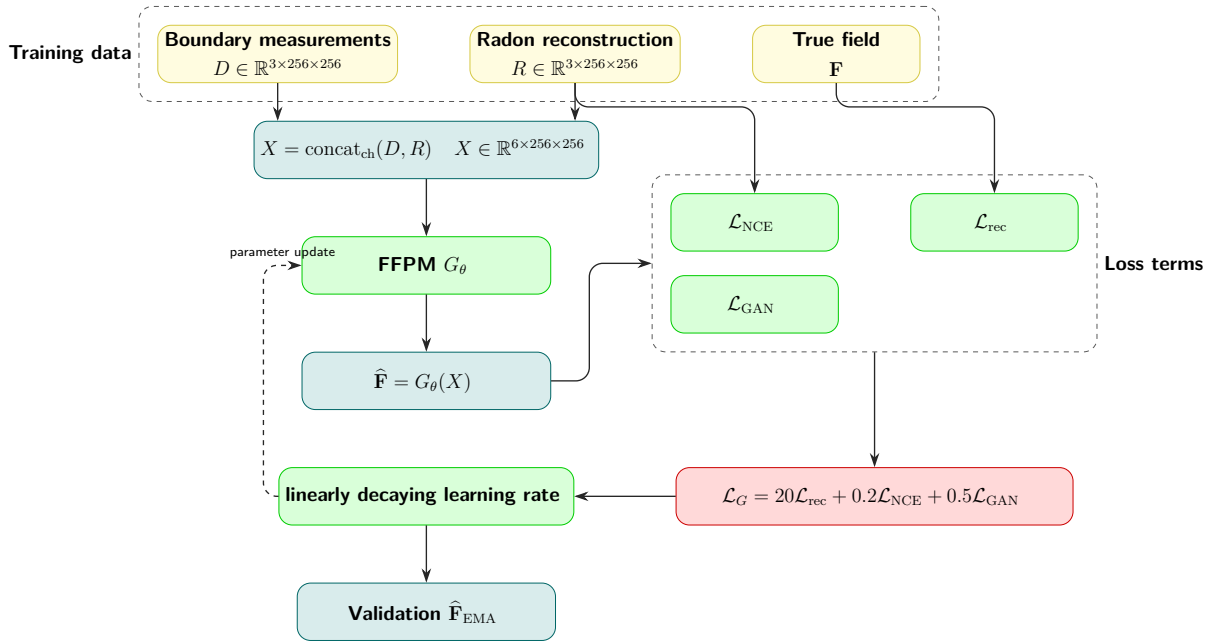

As shown in Fig.~\ref{fig:ml-training-evaluation-pipeline}, the basic per-pixel term \(\mathcal L_{\mathrm{rec}}= \|\widehat{\mathbf F}-\mathbf F\|_1\) is complemented by a two-scale \(\mathcal L_{\mathrm{GAN}}\) and a custom field-domain modification of NCE, motivated by PatchNCE \cite{oord2018cpc,park2020cut}. The latter maps local features \(\widehat{\mathbf F}\) and the Radon reconstruction at the same spatial positions across four scales. Therefore, the prior serves as a soft structural guide, but its pixel-by-pixel replication is not required; the precise definition of the term is given in the appendix~\ref{app:field-nce}.

The GAN term is activated after ten warm-up epochs. Each specialist model is trained for 150 epochs with a batch size of 16 using the Adam optimizer, \(\mathrm{lr}_G=4\cdot10^{-5}\), \(\mathrm{lr}_D=5\cdot10^{-5}\), and \((\beta_1,\beta_2)=(0.5,0.999)\). After the first half of training, the learning rate decreases linearly; automatic mixed precision and EMA of the generator weights are used.

\subsection{Informativeness of input data}\label{sec:ml-input-information}

To determine which part of the boundary data encodes the energy scale, we computed the spatial mean and standard deviation of each channel \(D_i\), defined in \eqref{eq:measurement-channels}, for every held-out sample and evaluated their Pearson correlations with the logarithm of the potential-field amplitude.

The strongest correlation was found for the time channel:
\begin{equation*}
\rho\!\left(
\left\langle D_3\right\rangle,
\log_{10}A
\right)
=
-0.909,
\qquad
\rho\!\left(
\sigma(D_3),
\log_{10}A
\right)
=
-0.681.
\end{equation*}
The negative sign corresponds to a decrease in the residence time of particles within the region with increasing field amplitude.

For channels \(D_1\) and \(D_2\), which contain information about the direction and magnitude of the final velocity, respectively, similar global statistics demonstrate a significantly weaker dependence:
\begin{align*}
\rho\!\left(
\left\langle D_1\right\rangle,
\log_{10}A
\right)
=
0.016,
\qquad
\rho\!\left(
\sigma(D_1),
\log_{10}A
\right)
=
0.025,
\\[1ex]
\rho\!\left(
\left\langle D_2\right\rangle,
\log_{10}A
\right)
=
0.014,
\qquad
\rho\!\left(
\sigma(D_2),
\log_{10}A
\right)
=
0.021.
\end{align*}

Thus, among the statistics considered, the time channel \(D_3\) demonstrates the most pronounced relationship with the potential field amplitude.

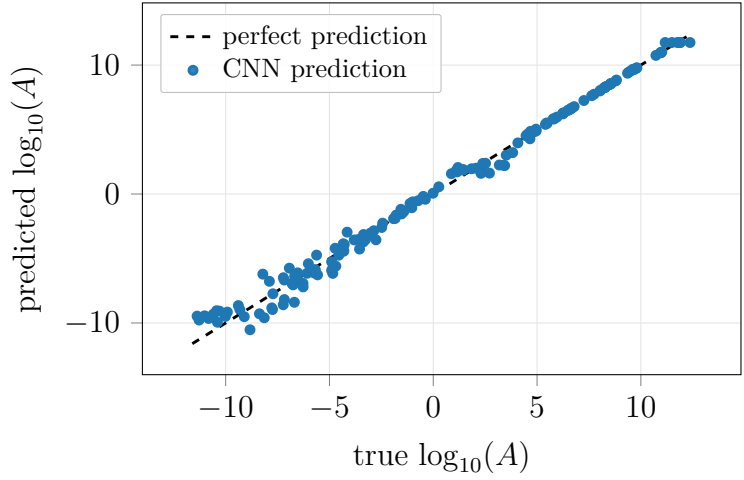
\begin{figure}[H]
\centering
\definecolor{scaleblack}{rgb}{0,0,0}
\definecolor{scaleblue}{rgb}{0.1216,0.4667,0.7059}
\begin{tikzpicture}
\begin{axis}[
  width=9.5cm,
  height=6.5cm,
  grid=both,
  grid style={line width=.3pt,draw=gray!22},
  tick align=outside,
  tick pos=left,
  every axis plot/.append style={line width=1.1pt,mark size=1.5pt},
  legend cell align=left,
  legend style={font=\footnotesize,draw=gray!50,fill=white,
    fill opacity=0.9,text opacity=1,rounded corners=1pt},
  xlabel={true $\log_{10}(A)$},
  ylabel={predicted $\log_{10}(A)$},
  legend pos=north west
]
\addplot[color=scaleblack,mark=none,dashed]
  coordinates {(-11.6079,-11.6079) (12.4198,12.4198)};
\addlegendentry{perfect prediction}
\addplot[color=scaleblue,mark=*,only marks] coordinates {
(6.61663,6.58174) (-5.74357,-6.11477) (9.70798,9.6865)
(3.43849,2.20496) (-6.7301,-6.93005) (-2.86476,-2.8475)
(-9.91777,-9.15671) (-9.31617,-8.96227) (-5.62073,-4.73764)
(3.18042,2.25052) (-6.93254,-5.74907) (-7.7789,-8.83216)
(-8.82728,-10.5268) (10.9463,10.9682) (2.38525,2.37426)
(-7.17652,-6.67034) (-5.66551,-6.03937) (-1.4217,-1.36724)
(-4.28818,-3.92385) (-4.72494,-4.20773) (6.68672,6.67204)
(6.63016,6.63269) (6.53283,6.51989) (-1.08257,-1.00802)
(0.873212,1.57553) (-3.60936,-3.52699) (4.67606,4.66918)
(5.45602,5.50652) (-5.6872,-5.85651) (1.84422,1.95907)
(-7.89386,-6.76892) (1.15201,1.71927) (-1.55803,-1.18418)
(4.90288,4.8735) (-7.22727,-8.58859) (8.55108,8.57239)
(4.96053,4.89968) (-6.52739,-6.11652) (-11.2811,-9.77327)
(-7.72349,-7.7373) (11.8973,11.7505) (4.07554,3.9651)
(-8.20966,-6.21186) (5.81846,5.8195) (1.45035,1.84974)
(-11.3769,-9.47083) (10.7255,10.7514) (9.50396,9.53364)
(-10.439,-9.05138) (1.1891,2.05524) (11.8167,11.7505)
(7.63199,7.6248) (0.266036,0.559072) (-4.54869,-4.74277)
(8.7924,8.78054) (8.26663,8.24867) (-10.3018,-9.08194)
(-10.3971,-9.94224) (-10.016,-9.49328) (6.47849,6.48084)
(8.82556,8.83901) (-4.14909,-2.95733) (9.77962,9.75534)
(-2.48464,-2.59543) (-1.83672,-1.89484) (2.3001,1.60232)
(-3.36285,-3.7134) (8.3189,8.33689) (-11.0151,-9.44536)
(-9.38979,-8.64911) (-4.69401,-5.59409) (-1.52485,-1.54018)
(-5.57333,-6.26829) (-0.953033,-0.618858) (3.38661,2.22075)
(4.46915,4.51898) (8.56051,8.54485) (-6.65268,-6.46459)
(-4.84068,-6.15397) (-10.8251,-9.6509) (11.764,11.7505)
(-6.7974,-6.95111) (2.33447,1.87648) (-1.00653,-0.621658)
(5.77509,5.83107) (3.51551,3.01634) (-6.27345,-6.92926)
(-3.26225,-3.53444) (11.4932,11.7505) (-6.0804,-6.14769)
(-6.01702,-5.40304) (5.48138,5.47839) (10.9738,10.9549)
(-0.383333,-0.408116) (-0.721363,-0.514584) (7.74605,7.73987)
(8.02729,7.9939) (-4.30673,-4.43265) (9.36002,9.36774)
(6.25025,6.21568) (8.23166,8.24423) (-7.22139,-6.49047)
(-1.11895,-0.737432) (-3.35722,-3.14071) (-3.55158,-4.26827)
(-6.75138,-7.04648) (11.0064,11.0123) (-7.74384,-8.95816)
(-8.37296,-9.28721) (9.81033,9.79162) (-4.8978,-5.92334)
(8.49498,8.48526) (-7.17414,-8.18238) (4.80047,4.83985)
(11.9006,11.7505) (5.95874,5.96754) (-2.43667,-2.2616)
(8.05951,8.04693) (7.25672,7.2524) (-6.55564,-6.43269)
(-0.0110633,0.0614279) (1.43497,1.90246) (2.05999,2.00373)
(-4.82806,-5.68416) (-1.9124,-1.91188) (8.54028,8.537)
(-4.33574,-3.85788) (12.3676,11.7505) (9.60795,9.66893)
(-2.76002,-3.55219) (2.51821,2.3842) (6.27739,6.28639)
(-6.52601,-6.64035) (-6.68942,-8.39627) (5.40381,5.4029)
(-3.8078,-3.56101) (4.79042,4.77919) (-6.26929,-7.20191)
(9.41418,9.44853) (6.23578,6.26402) (4.94509,5.03056)
(-3.02401,-3.03092) (-4.89689,-5.25289) (-10.5976,-9.3211)
(6.78624,6.77773) (8.28552,8.2863) (2.70777,1.62095)
(-0.481587,-0.185686) (-8.14143,-9.59667) (-1.03041,-1.06215)
(1.10803,1.82248) (-1.78451,-1.63004) (-6.67684,-6.35023)
(-9.10496,-9.51442) (3.82503,3.2103) (4.53297,4.62992)
(9.67158,9.66171) (11.1712,11.7505) (4.65675,4.28181)
(4.68072,4.8594)
};
\addlegendentry{CNN prediction}
\end{axis}
\end{tikzpicture}
\caption{
Prediction of \(\log_{10}A\) by the auxiliary Img2Vec regressor from the measurement tensor \(D\), without using the Radon reconstruction. The dashed diagonal denotes exact prediction.
}

\label{fig:ml-scale-regression}
\end{figure}

To test the feasibility of nonlinear amplitude scale extraction, an auxiliary Img2Vec model was additionally trained, predicting \(\log_{10}A\) from the full measurement tensor \(D\). Radon reconstruction \(R\) was not used, since the goal of the experiment was to assess the information content of directly measured data. The obtained predictions are mainly located near the diagonal of the exact reconstruction, see Fig.~\ref{fig:ml-scale-regression}. Thus, the tensor \(D\) allows to reconstruct the amplitude scale, while the dominant role of the time channel \(D_3\) is established by the above correlation analysis.

The auxiliary Img2Vec model was trained on a separate dataset, covering the range
\begin{equation*}
A\in[10^{-12},10^{12}],
\qquad
\log_{10}A\in[-12,12].
\end{equation*}
This range is wider than the set of amplitudes used to train the specialized FFPM models.

\subsection{Applicability limit}\label{sec:ml-applicability}

\subsubsection{Energy parameterization}\label{sec:ml-energy}

To compare reconstruction quality across field amplitudes, we introduce the ratio of the initial kinetic energy to a characteristic potential-energy scale:
\begin{equation}
\kappa
=
\frac{E_{\mathrm{kin}}}{E_{\mathrm{pot}}},
\qquad
E_{\mathrm{kin}}
=
\frac{v_0^2}{2}.
\label{eq:energy-ratio}
\end{equation}

This parametrization can be related to the perturbative parameter introduced in Section~\ref{subsec:finite-speed-expansion}. Using \(E_{\mathrm{pot}}\) as an empirical estimate of the characteristic potential variation \(\delta U\), we obtain
\begin{equation}
\epsilon
=
\frac{\delta U}{v_0^2}
=
\frac{E_{\mathrm{pot}}}{2E_{\mathrm{kin}}}
=
\frac{1}{2\kappa}\text{ for $m=1$}.
\label{eq:epsilon-eta}
\end{equation}
Thus, the high-energy limit \(\epsilon\to0\) corresponds to \(\kappa\to\infty\), whereas decreasing \(\kappa\) corresponds to increasing nonlinearity and approaching the theoretically predicted loss of recoverability.

In this case, the set of potentials must be characterized by some energy characteristic of this set related to its shape:
\begin{equation*}
E_{\mathrm{pot}}^{(k)}=C_kA,
\qquad
\kappa_k=\frac{v_0^2}{2C_kA}.
\end{equation*}

We consider five physically motivated definitions of this characteristic scale:
\begin{align}
E_{\mathrm{pot}}^{(1)}
&=
\left\langle |U|\right\rangle_{\Omega},
&
C_1&=0.551,
\notag
\\
E_{\mathrm{pot}}^{(2)}
&=
\max_{\mathbf x\in\Omega}|U(\mathbf x)|,
&
C_2&=0.973,
\notag
\\
E_{\mathrm{pot}}^{(3)}
&=
\max_{\mathbf x\in\partial\Omega}|U(\mathbf x)|
-
\left\langle |U|\right\rangle_{\partial\Omega},
&
C_3&=0.272,
\notag
\\
E_{\mathrm{pot}}^{(4)}
&=
\max_{\mathbf x\in\Omega}|U(\mathbf x)|
-
\left\langle |U|\right\rangle_{\partial\Omega},
&
C_4&=0.497,
\label{eq:energy-scale-c4}
\\
E_{\mathrm{pot}}^{(5)}
&=
\max_{\mathbf x\in\Omega}|U(\mathbf x)|
-
\max_{\mathbf x\in\partial\Omega}|U(\mathbf x)|,
&
C_5&=0.225.
\notag
\end{align}

It is also interesting to compare this with the naive definition of such a recovery boundary. We introduce a local excess of potential over its characteristic boundary level:
\begin{equation}
\Delta U(\mathbf x)
=
|U(\mathbf x)|
-
\left\langle |U|\right\rangle_{\partial\Omega}.
\label{eq:local-potential-excess}
\end{equation}

In the naive energy approximation, a point is considered attainable if the initial kinetic energy exceeds the corresponding local change in potential defined in \eqref{eq:local-potential-excess}:
\begin{align}
f_{\mathrm{acc}}
&=
\frac{
\left|
\left\{
\mathbf x\in\Omega_{0.85}:
\Delta U(\mathbf x)\leq E_{\mathrm{kin}}
\right\}
\right|
}{
\left|\Omega_{0.85}\right|
},
\notag
\\
f_{\mathrm{inacc}}
&=
\frac{
\left|
\left\{
\mathbf x\in\Omega_{0.85}:
\Delta U(\mathbf x)>E_{\mathrm{kin}}
\right\}
\right|
}{
\left|\Omega_{0.85}\right|
}
=
1-f_{\mathrm{acc}}.
\label{eq:naive-accessibility}
\end{align}

This definition is a local energy indicator and does not take into account the connectivity of the set of reachable points. In particular, a region satisfying the local energy condition may be separated from the boundary by an energetically inaccessible barrier. Such topological restrictions are intentionally ignored here, since \(f_{\mathrm{inacc}}\) is used only as a simple diagnostic characteristic, not as a strict measure of dynamic reachability.

To explore the applicability of the method, specialist models were independently trained for the 16 amplitudes defined in \eqref{eq:ffpm-amplitude-grid}. The same field families, architecture, training set size, and estimation procedure were used in all cases. Each point of the final curve was calculated using 200 samples from a fixed validation portion of the corresponding dataset.

Fig.~\ref{fig:ml-mix-v2-24pt-multimetric} shows the resulting dependences of the errors on the energy ratio in \eqref{eq:energy-ratio} for different energy characteristics. On a logarithmic scale, differences in $C$ appear as a horizontal shift. The dashed line shows the naively inaccessible area fraction \(f_{\mathrm{inacc}}\) defined in \eqref{eq:naive-accessibility}.

\input{mix_v2_24pt_multimetric.tex}

The scale in \eqref{eq:energy-scale-c4} is used below because it measures the characteristic potential variation between the interior of the domain and its boundary. The resulting dependence is shown separately in Fig.~\ref{fig:ml-mix-v2-red-fit-inaccessible}. A pronounced increase in the reconstruction error is observed for \(\kappa\in[1,3]\). Using \eqref{eq:epsilon-eta}, this corresponds to
\begin{equation*}
\epsilon=\frac{1}{2\kappa}\in
\left[\frac{1}{6},\frac{1}{2}\right].
\end{equation*}
This interval includes the finite-energy regime in which the theoretical analysis predicts increasing nonlinear corrections and, for the harmonic example in Section~\ref{sec:hp}, the critical value \(\epsilon_c=1/2\).

\input{mix_v2_red_fit_inaccessible.tex}

\subsubsection{Out-of-distribution test}\label{sec:ml-ood}

To evaluate generalization beyond the training distribution, we use a centered two-dimensional paraboloid. Potentials of this form were not included in the training set, making this an analytical out-of-distribution example. As in the main evaluation protocol, FFPM predicts the force field, and the potential is subsequently obtained during postprocessing.

\begin{figure}[ht!]
\centering
\definecolor{parradon}{rgb}{0.1216,0.4667,0.7059}
\definecolor{parffpm}{rgb}{1.0000,0.4980,0.0549}
\definecolor{parvalid}{rgb}{0.1725,0.6275,0.1725}
\resizebox{\linewidth}{!}{%
\begin{tikzpicture}
\begin{groupplot}[
  group style={group size=2 by 1,horizontal sep=1.4cm},
  width=8.0cm,
  height=6.3cm,
  grid=both,
  grid style={line width=.3pt,draw=gray!22},
  tick align=outside,
  tick pos=left,
  every axis plot/.append style={line width=1.1pt,mark size=1.8pt},
  xmode=log,
  x dir=reverse,
  xlabel={$E_{\mathrm{kin}}/\Delta V$
    ($\leftarrow$ weak field \textbar{} strong field $\rightarrow$)},
  legend cell align=left,
  legend style={
    font=\scriptsize,
    draw=gray!50,
    fill=white,
    fill opacity=0.94,
    text opacity=1,
    rounded corners=1pt
  }
]

\nextgroupplot[
  title={(a) Amplitude-sensitive error},
  ylabel={potential relMAE, DC-aligned, $r<0.85$},
  ymin=0,
  ymax=1.03,
  legend style={at={(0.5,-0.30)},anchor=north,legend columns=1}
]
\addplot[color=parradon,mark=square*] coordinates {
  (1280,0.124651) (640,0.123896) (320,0.122572) (160,0.119985)
  (80,0.114811) (40,0.104589) (20,0.0849052) (10,0.0505888)
  (5,0.0681293) (2.5,0.202408) (1.25,0.407102) (0.625,0.582695)
  (0.3125,0.728544) (0.15625,0.843732) (0.078125,0.915962)
  (0.0390625,0.955971)
};
\addlegendentry{Radon reconstruction $\rightarrow\widehat U$}
\addplot[color=parffpm,mark=*] coordinates {
  (1280,0.0703135) (640,0.056909) (320,0.0547667) (160,0.0348057)
  (80,0.0499553) (40,0.0470328) (20,0.0653013) (10,0.104173)
  (5,0.142681) (2.5,0.236036) (1.25,0.350861) (0.625,0.425178)
  (0.3125,0.559597) (0.15625,0.594381) (0.078125,0.799662)
  (0.0390625,0.59134)
};
\addlegendentry{FFPM: $\widehat{\mathbf F}\rightarrow\widehat U$}
\draw[gray,densely dotted]
  ({axis cs:1,0}|-{rel axis cs:0,0}) --
  ({axis cs:1,0}|-{rel axis cs:0,1});

\nextgroupplot[
  title={(b) Shape and trajectory diagnostics},
  ylabel={correlation / valid-ray fraction},
  yticklabel pos=right,
  ylabel style={at={(axis description cs:1.17,0.5)},anchor=south},
  ymin=0.74,
  ymax=1.01,
  legend style={at={(0.5,-0.30)},anchor=north,legend columns=1}
]
\addplot[color=parradon,mark=square*] coordinates {
  (1280,0.999885) (640,0.999882) (320,0.999876) (160,0.999863)
  (80,0.999837) (40,0.999776) (20,0.999627) (10,0.999244)
  (5,0.998254) (2.5,0.996136) (1.25,0.993516) (0.625,0.980867)
  (0.3125,0.98189) (0.15625,0.987388) (0.078125,0.989867)
  (0.0390625,0.989661)
};
\addlegendentry{Radon potential correlation}
\addplot[color=parffpm,mark=*] coordinates {
  (1280,0.997025) (640,0.999641) (320,0.999442) (160,0.999367)
  (80,0.999113) (40,0.998698) (20,0.997615) (10,0.994007)
  (5,0.987706) (2.5,0.968838) (1.25,0.940375) (0.625,0.94787)
  (0.3125,0.888554) (0.15625,0.891897) (0.078125,0.890479)
  (0.0390625,0.785144)
};
\addlegendentry{FFPM potential correlation}

\end{groupplot}
\end{tikzpicture}%
}
\caption{ Out-of-distribution reconstruction of a centered two-dimensional paraboloid. }
\label{fig:ml-paraboloid-applicability}
\end{figure}
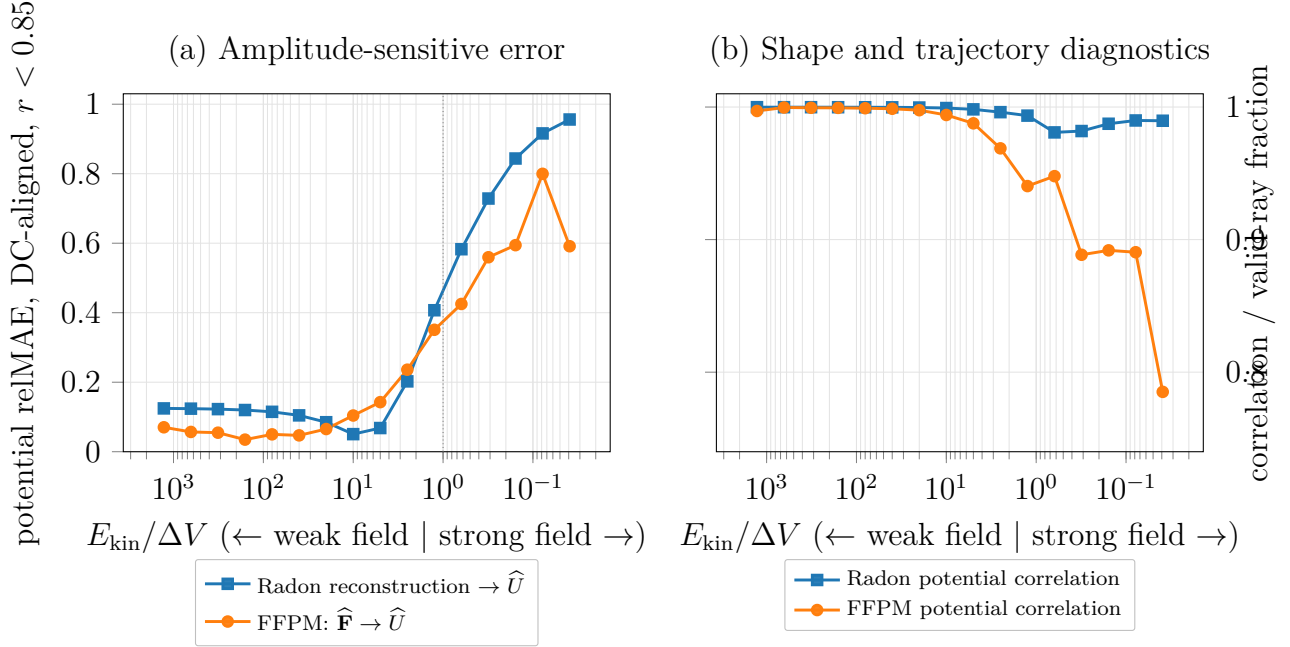

The energy dependence of the error on the paraboloid is shown in Fig.~\ref{fig:ml-paraboloid-applicability}. It is evident that going beyond the training distribution is not accompanied by a separate sharp jump in the error: the deterioration begins in the same energy region as in the main set. This is consistent with the interpretation of the transition as a consequence of a decrease in the information content of the forward operator in a strong field, and not just as a typical model generalization error.

The two methods demonstrate qualitatively different degradation mechanisms. At low \(E_{\mathrm{kin}}/\Delta V\), Radon reconstruction retains very high correlation but loses amplitude and approaches a nearly flat potential. Correlation is weakly sensitive to such multiplicative gain, so it masks the physically significant loss of dynamic range. FFPM, on the other hand, retains a pronounced potential well for longer, but in strong fields it acquires noise, anisotropic deformations, and radial symmetry breaking.

\begin{figure}[ht!]
\centering
\includegraphics[
width=\linewidth,
trim={0cm 0cm 0cm 0.9cm},
clip
]{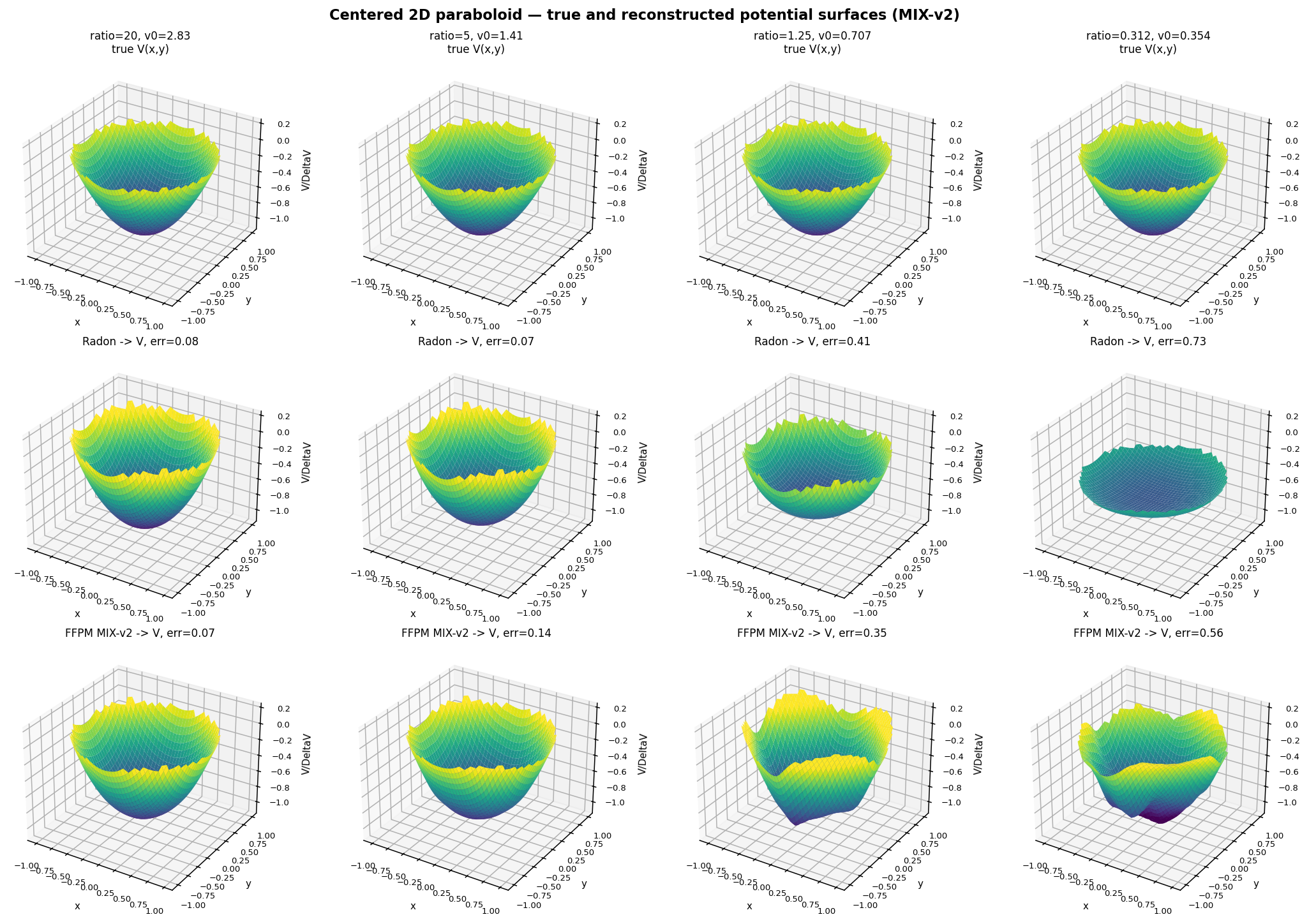}
\caption{
True, Radon-reconstructed, and FFPM-reconstructed centered paraboloid potentials for decreasing \(E_{\mathrm{kin}}/\Delta V\).
}

\label{fig:ml-paraboloid-gallery}
\end{figure}

Fig.~\ref{fig:ml-paraboloid-gallery} further illustrates these distinct failure modes. The Radon reconstruction preserves the axial symmetry of the paraboloid but progressively loses its shape and amplitude. FFPM retains a central minimum and elevated boundary values, but introduces substantial noise and geometric distortions. Consequently, similar integral errors may correspond to qualitatively different reconstruction defects.

\section{Conclusions and discussion}\label{sec:concl}

We studied the inverse scattering problem for Newtonian particles at finite initial velocity. Exact calculations for constant-force and harmonic potentials show that the structure of the scattering map changes when the initial kinetic energy becomes comparable to a characteristic potential variation between the boundary and the interior of the domain. These examples motivate an energy-based description of the transition between the high-energy Radon regime and the strongly nonlinear regime. We also derived an iterative reconstruction procedure that organizes finite-velocity corrections as an expansion about the infinite-velocity limit.

For the numerical inverse problem, we combined a CUDA forward simulator, a physically motivated Radon approximation, and the Force-Field Prediction Model (FFPM). FFPM reconstructs the conservative force field \(\widehat{\mathbf F}\); when required, the corresponding potential \(\widehat U\) is obtained afterwards by least-squares integration and is defined up to an additive constant. Experiments on synthetic random fields provide a numerical characterization of the reconstruction error as a function of the kinetic-to-potential energy ratio. The same qualitative transition is observed for a centered paraboloid excluded from the training distribution, indicating that the degradation is not explained solely by ordinary out-of-distribution generalization error. 

The theoretical and numerical results are therefore consistent with a common energy-scale interpretation: reconstruction deteriorates when an increasing part of the potential landscape becomes weakly accessible to the available trajectories. The Radon approximation and FFPM exhibit qualitatively different failure modes. The former tends to lose the amplitude of the potential while preserving its coarse shape, whereas the latter may retain physically relevant extrema but introduce noise and geometric distortions. Consequently, integral metrics such as \(L_1\), \(L_2\), and correlation do not fully characterize the physical quality of a reconstruction.

Future work will address independent test sets and repeated training runs, robustness to noise and missing trajectories, and the incorporation of a differentiable forward operator into training. It is also natural to augment the evaluation with critical-point statistics, sublevel-set connectivity, and persistence-based distances \cite{edelsbrunner2002persistence,cohensteiner2007stability}. Further extensions include non-conservative or vortical force fields and particle dynamics with viscous or dry friction.

\section{Acknowledgments}

We are grateful to Alexey Yur'evich Morozov for his valuable comments on the text and discussions of related topics. We also thank Maxim Valerievich Fedorov, Dmitry Viktorovich Vasilyev for discussions on related topics.

This work was supported by The Ministry of Economic Development of the Russian Federation (IGK 000000C313925P4C0002), agreement No139-15-2025-010.

\newpage


\appendix

\section{Detailed reconstructor architecture}\label{app:reconstructor-architecture}

FFPM uses a Johnson-style \(6\rightarrow3\) generator consisting of an initial \(7\times7\) convolution, two downsampling blocks, six residual blocks operating at a spatial resolution of \(64\times64\), two upsampling blocks, and a final \(7\times7\) convolution followed by \(\tanh\). The generator contains \(8{,}137{,}539\) trainable parameters. No long encoder--decoder skip connections are used; residual connections are restricted to the bottleneck blocks.

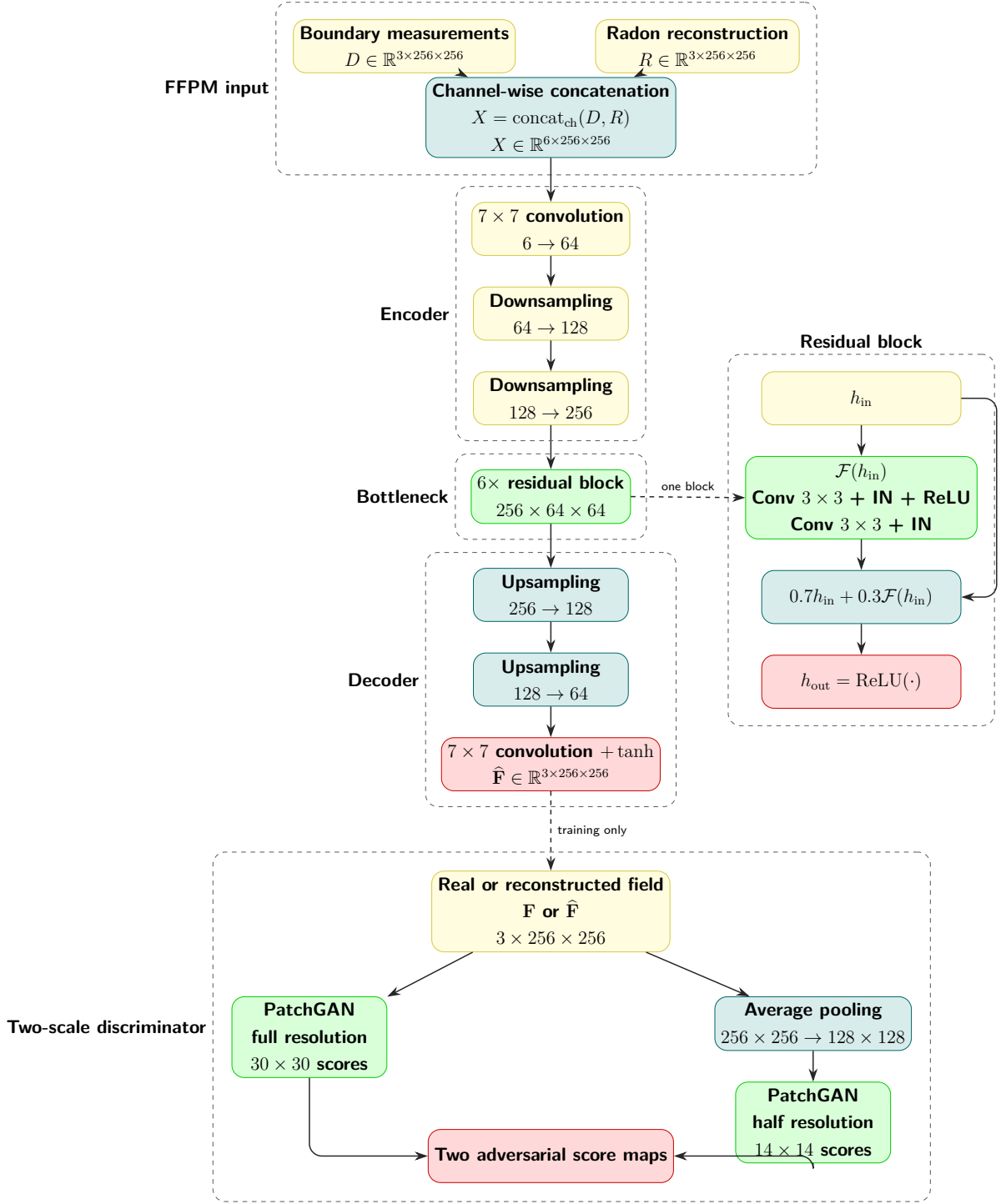
\begin{figure}[ht!]
\centering
\resizebox{0.93\linewidth}{!}{%
\begin{tikzpicture}[node distance=7mm and 22mm]

  \node (measurements) [modern=yellow, minimum width=4.6cm]
    {Boundary measurements\\
     $D\in\mathbb{R}^{3\times256\times256}$};

  \node (radon) [modern=yellow, right=18mm of measurements,
    minimum width=4.6cm]
    {Radon reconstruction\\
     $R\in\mathbb{R}^{3\times256\times256}$};

  \coordinate (inputmid) at ($(measurements)!0.5!(radon)$);

  \node (input) [modern=teal, below=of inputmid,
    minimum width=5.4cm]
    {Channel-wise concatenation\\
     $X=\operatorname{concat}_{\mathrm{ch}}(D,R)$\\
     $X\in\mathbb{R}^{6\times256\times256}$};

  \node (inputgroup) [
    draw=black!65, dashed, rounded corners=8pt,
    fit=(measurements)(radon)(input), inner sep=11pt,
    label={[font=\sffamily\bfseries]left:FFPM input}
  ] {};

  \node (conv) [modern=yellow, below=10mm of input]
    {$7\times7$ convolution\\$6\rightarrow64$};
  \node (down1) [modern=yellow, below=of conv]
    {Downsampling\\$64\rightarrow128$};
  \node (down2) [modern=yellow, below=of down1]
    {Downsampling\\$128\rightarrow256$};
  \node (res) [modern=green, below=10mm of down2]
    {$6\times$ residual block\\$256\times64\times64$};
  \node (up1) [modern=teal, below=10mm of res]
    {Upsampling\\$256\rightarrow128$};
  \node (up2) [modern=teal, below=of up1]
    {Upsampling\\$128\rightarrow64$};
  \node (output) [modern=red, below=of up2, minimum width=4.6cm]
    {$7\times7$ convolution $+\tanh$\\
     $\widehat{\mathbf F}\in\mathbb{R}^{3\times256\times256}$};

  \draw[arrowstyle] (measurements) -- (input);
  \draw[arrowstyle] (radon) -- (input);
  \draw[arrowstyle] (input) -- (conv);
  \draw[arrowstyle] (conv) -- (down1);
  \draw[arrowstyle] (down1) -- (down2);
  \draw[arrowstyle] (down2) -- (res);
  \draw[arrowstyle] (res) -- (up1);
  \draw[arrowstyle] (up1) -- (up2);
  \draw[arrowstyle] (up2) -- (output);

  \node (encoder) [
    draw=black!65, dashed, rounded corners=8pt,
    fit=(conv)(down1)(down2), inner sep=10pt,
    label={[font=\sffamily\bfseries]left:Encoder}
  ] {};
  \node (bottleneck) [
    draw=black!65, dashed, rounded corners=8pt,
    fit=(res), inner sep=10pt,
    label={[font=\sffamily\bfseries]left:Bottleneck}
  ] {};
  \node (decoder) [
    draw=black!65, dashed, rounded corners=8pt,
    fit=(up1)(up2)(output), inner sep=10pt,
    label={[font=\sffamily\bfseries]left:Decoder}
  ] {};

  \node (rin) [modern=yellow, right=30mm of down2,
    minimum width=4.5cm] {$h_{\mathrm{in}}$};
  \node (rfun) [modern=green, below=of rin,
    minimum width=4.5cm]
    {$\mathcal F(h_{\mathrm{in}})$\\
     Conv $3\times3$ + IN + ReLU\\
     Conv $3\times3$ + IN};
  \node (rmix) [modern=teal, below=of rfun,
    minimum width=4.5cm]
    {$0.7h_{\mathrm{in}}+0.3\mathcal F(h_{\mathrm{in}})$};
  \node (rout) [modern=red, below=of rmix,
    minimum width=4.5cm]
    {$h_{\mathrm{out}}=\operatorname{ReLU}(\cdot)$};

  \draw[arrowstyle] (rin) -- (rfun);
  \draw[arrowstyle] (rfun) -- (rmix);
  \draw[arrowstyle] (rmix) -- (rout);
  \draw[arrowstyle] (rin.east) -- ++(8mm,0) |- (rmix.east);
  \draw[arrowstyle,dashed] (res.east) --
    node[above,font=\sffamily\scriptsize]{one block} (rfun.west);

  \node (resdetail) [
    draw=black!65, dashed, rounded corners=8pt,
    fit=(rin)(rfun)(rmix)(rout), inner sep=11pt,
    label={[font=\sffamily\bfseries]above:Residual block}
  ] {};

  \node (discinput) [modern=yellow, below=18mm of output,
    minimum width=4.8cm]
    {Real or reconstructed field\\
     $\mathbf F$ or $\widehat{\mathbf F}$\\
     $3\times256\times256$};
  \node (discfull) [modern=green,
    below left=10mm and 10mm of discinput]
    {PatchGAN\\full resolution\\$30\times30$ scores};
  \node (pool) [modern=teal,
    below right=10mm and 10mm of discinput]
    {Average pooling\\$256\times256\rightarrow128\times128$};
  \node (dischalf) [modern=green, below=of pool]
    {PatchGAN\\half resolution\\$14\times14$ scores};
  \node (scores) [modern=red, below=13mm of discinput,
    yshift=-27mm, minimum width=4.4cm]
    {Two adversarial score maps};

  \draw[arrowstyle,dashed] (output) --
    node[right,font=\sffamily\scriptsize]{training only} (discinput);
  \draw[arrowstyle] (discinput) -- (discfull);
  \draw[arrowstyle] (discinput) -- (pool);
  \draw[arrowstyle] (pool) -- (dischalf);
  \draw[arrowstyle] (discfull.south) |- (scores.west);
  \draw[arrowstyle] (dischalf.south) |- (scores.east);

  \node (discriminator) [
    draw=black!65, dashed, rounded corners=8pt,
    fit=(discinput)(discfull)(pool)(dischalf)(scores), inner sep=12pt,
    label={[font=\sffamily\bfseries]left:Two-scale discriminator}
  ] {};

\end{tikzpicture}%
}
\caption{Detailed FFPM architecture. The generator follows an encoder--residual-bottleneck--decoder design without long skip connections. A single residual block is detailed on the right. Two unconditional PatchGAN discriminators operate at the original and downsampled resolutions and are used only during training.}
\label{fig:ml-model-architecture}
\end{figure}

A residual block implements the transformation
\begin{equation}
h_{\mathrm{out}}
=
\operatorname{ReLU}
\left[
0.7h_{\mathrm{in}}
+
0.3\mathcal F(h_{\mathrm{in}})
\right],
\end{equation}
where \(\mathcal F\) consists of two \(3\times3\) convolutions with Reflection Padding and Instance Normalization; ReLU is applied after the first convolution. The encoder changes the spatial resolution according to \(256^2\rightarrow128^2\rightarrow64^2\), and the decoder restores it using two transposed convolutions. The output channels are ordered as \((0,F_y,F_x)\).

Two unconditional PatchGAN discriminators are used during training. The first operates at the original resolution and produces a \(30\times30\) score map with a \(70\times70\) receptive field. The second receives the field after \(2\times2\) average pooling and produces a \(14\times14\) score map. Its receptive field with respect to the original image is \(140\times140\). The discriminators receive only \(\mathbf F\) or \(\widehat{\mathbf F}\); they do not receive the boundary measurements and are not used at inference time.

\subsection{Contrastive loss term}\label{app:field-nce}

Let \(E_l\) denote level \(l\) of the shared field encoder, \(M_l\) the corresponding projection MLP, and \(s_l(p)\) a selected spatial position. The normalized prediction and Radon reconstruction features are defined as 

\begin{align}
q_{b,p}^{(l)}
&=
\operatorname{norm}_2
\left[
M_l\!\left(E_l(\widehat{\mathbf F}_b)_{s_l(p)}\right)
\right],
\\
k_{b,p}^{(l)}
&=
\operatorname{sg}
\left\{
\operatorname{norm}_2
\left[
M_l\!\left(E_l(R_b)_{s_l(p)}\right)
\right]
\right\},
\end{align}

where \(\operatorname{sg}\) stops the gradient through the prior branch. After combining the sample and spatial-position indices into a single index \(i\), the loss at level \(l\) is 

\begin{equation}
\mathcal L_l
=
-\frac{1}{N_l}
\sum_{i=1}^{N_l}
\log
\frac{\exp(q_i^{T}k_i/\tau)}
{\exp(q_i^{T}k_i/\tau)
+\sum_{j\ne i}\exp(q_i^{T}k_j/\tau)},
\qquad
\tau=0.2,
\end{equation}
and the full contrastive term is
\begin{equation}
\mathcal L_{\mathrm{NCE}}
=
\frac14\sum_{l=0}^{3}\mathcal L_l.
\end{equation}
At each of the four levels, 128 shared spatial positions are sampled; the remaining positions in the current batch form the negative pairs.

\section{Mathematical definitions of synthetic potential fields}\label{app:synthetic-fields}

The potential-field families listed in Section~\ref{sec:ml-synthetic-fields} are defined below. The common postprocessing pipeline applied to all realizations is described at the end of this appendix. 
\subsection{Gaussian fields with RBF covariance}\label{app:field-rbf}

The field is defined as a Gaussian process
\begin{equation}
U_{\mathrm{RBF}}(x)
\sim
\mathcal{GP}\!\left(m(x),k_{\mathrm{RBF}}(x,x')\right),
\qquad
k_{\mathrm{RBF}}(x,x')
=
\sigma^2
\exp\left[
-\frac12
(x-x')^{T}\Lambda^{-1}(x-x')
\right],
\end{equation}
where \(\Lambda=\operatorname{diag}(\ell_1^2,\ldots,\ell_d^2)\). This covariance produces smooth realizations possessing mean-square derivatives of all orders \cite{RasmussenWilliams2006,Kanagawa2018}.

\subsection{Matern fields}\label{app:field-matern}

A Matern field is defined as

\begin{equation}
U_{\mathrm{Mat}}(x)
\sim
\mathcal{GP}\!\left(m(x),k_\nu(x,x')\right),
\end{equation}
where, for \(r=\lVert x-x'\rVert\),
\begin{equation}
k_\nu(r)
=
\sigma^2
\frac{2^{1-\nu}}{\Gamma(\nu)}
\left(
\frac{\sqrt{2\nu}\,r}{\ell}
\right)^\nu
K_\nu\left(
\frac{\sqrt{2\nu}\,r}{\ell}
\right).
\end{equation}
Here \(\nu>0\) controls the smoothness of the realization, \(\ell>0\) sets the characteristic correlation length, and \(K_\nu\) denotes the modified Bessel function of the second kind \cite{GuttorpGneiting2006,Lindgren2011,Lindgren2022}.

\subsection{Truncated Fourier series}\label{app:field-fourier}

The potential is represented by a finite random sum of harmonics:
\begin{equation}
U_{\mathrm{Fourier}}(x)
=
\sum_{m=1}^{T}
\frac{\xi_m}
{\bigl(\lVert k_m\rVert^2+1\bigr)^{s/2}}
\sin\left(2\pi k_m^{T}x+\phi_m\right),
\qquad
T\sim\operatorname{Unif}\{10,\ldots,50\}.
\end{equation}
Here \(k_m\), \(\phi_m\), and \(\xi_m\) denote the wave vector, random phase, and amplitude coefficient, respectively, while \(s\) controls the spectral decay and hence the smoothness of the field \cite{ShinozukaDeodatis1996,Kurbanmuradov2013,Chepurnov2021}.

\subsection{Signed sums of Gaussian basis functions}\label{app:field-gmm}

The potential is defined as a signed sum of Gaussian basis functions:
\begin{equation}
U_{\mathrm{GMM}}(x)
=
\sum_{j=1}^{M}
a_j
\exp\left[
-\frac{1}{2}
\left(
\frac{(x_1-c_{j1})^2}{s_{j1}^2}
+
\frac{(x_2-c_{j2})^2}{s_{j2}^2}
\right)
\right],
\qquad
a_j\in\mathbb R.
\end{equation}
The centers \(c_j\) and scales \(s_{j1},s_{j2}\) define anisotropic, axis-aligned components. Unlike a probabilistic Gaussian mixture, the coefficients \(a_j\) may have either sign and are not constrained to sum to one \cite{Green2018,McLachlan2019}.

\subsection{Perlin noise}\label{app:field-perlin}

For \(x\in\mathbb R^d\), define
\begin{equation}
n=\lfloor x\rfloor,
\qquad
t=x-\lfloor x\rfloor\in[0,1)^d,
\qquad
s(u)=6u^5-15u^4+10u^3.
\end{equation}
Each lattice vertex \(n+v\), where \(v\in\{0,1\}^d\), is assigned a pseudorandom unit gradient \(g_{n+v}\). The interpolation weights and noise values are defined by
\begin{align}
w_v(t)
&=
\prod_{j=1}^{d}
\left[
v_j s(t_j)
+
(1-v_j)\bigl(1-s(t_j)\bigr)
\right],
\\
N_{\mathrm{Perlin}}(x)
&=
\sum_{v\in\{0,1\}^d}
w_v(t)
\left\langle
g_{n+v},t-v
\right\rangle.
\end{align}
The multiscale potential is constructed as a sum of octaves:
\begin{equation}
U_{\mathrm{Perlin}}(x)
=
\sum_{q=0}^{Q-1}
a_0\rho^q
N_{\mathrm{Perlin}}\!\left(\lambda^q x\right),
\qquad
\lambda>1,
\qquad
0<\rho<1.
\end{equation}
The parameters \(\lambda>1\) and \(0<\rho<1\) set, respectively, the increase in spatial frequency and the decay in octave amplitude \cite{Perlin1985,Perlin2002,Lagae2010,Jain2022}.

\subsection{Sums of compact harmonic wells}\label{app:field-compact}

For the local coordinates
\begin{equation}
\begin{pmatrix}
u_j(x)\\
v_j(x)
\end{pmatrix}
=
\begin{pmatrix}
\cos\theta_j & \sin\theta_j\\
-\sin\theta_j & \cos\theta_j
\end{pmatrix}
(x-c_j),
\end{equation}
define
\begin{equation}
q_j(x)
=
\frac{u_j(x)^2}{r_{x,j}^2}
+
\frac{v_j(x)^2}{r_{y,j}^2}.
\end{equation}
The total potential is
\begin{equation}
U_{\mathrm{comp}}(x)
=
-\sum_{j=1}^{M}
a_j\bigl(1-q_j(x)\bigr)_+,
\qquad
(z)_+=\max(z,0).
\end{equation}
Each component is quadratic inside the ellipse \(q_j<1\) and vanishes outside it. The potential is \(C^0\), but generally not \(C^1\), across the boundary of its support \cite{Wendland1995,Lewis1989,Lagae2009}.

\subsection{Sums of anisotropic Gaussian wells}\label{app:field-anisotropic}

The number of components and their centers are sampled according to
\begin{equation}
K\sim\operatorname{Unif}\{1,\ldots,16\},
\qquad
c_j\in D_{0.78},
\qquad
D_{0.78}
=
\left\{
x\in\mathbb R^2:\lVert x\rVert\leq0.78
\right\}.
\end{equation}
The shape matrix of component \(j\) is
\begin{equation}
\Sigma_j
=
Q_j
\begin{pmatrix}
\ell_{j,1}^2 & 0\\
0 & \ell_{j,2}^2
\end{pmatrix}
Q_j^{T},
\end{equation}
where \(Q_j\) determines the orientation of the principal axes, while \(\ell_{j,1}\) and \(\ell_{j,2}\) are their spatial scales.

The normalized sum of anisotropic Gaussian wells is defined as
\begin{equation}
U_{\mathrm{AG}}(x)
=
-
\frac{
\displaystyle
\sum_{j=1}^{K}
a_j
\exp\left[
-\frac{1}{2}
(x-c_j)^{T}
\Sigma_j^{-1}
(x-c_j)
\right]
}{
\displaystyle
\sum_{j=1}^{K}a_j
},
\qquad
a_j>0.
\end{equation}
The normalization reduces the dependence of the overall potential scale on the number of components and on the absolute values of the coefficients \(a_j\) \cite{Kormann2019,Beatson2010,Casciola2006}.

\subsection{Common processing pipeline for the mixed conservative-field dataset}\label{app:field-processing}

For each generator, the raw potential \(U_{\mathrm{raw}}\) is centered and normalized by its standard deviation over the computational grid. A nonlinear amplitude transformation, the boundary-condition mask, the disk mask, and the final amplitude factor are then applied:
\begin{equation}
U(\mathbf{x})
=
A\,M_{\mathrm{BC}}(x)M_{\mathrm{disk}}(x)
\tanh\left[
\alpha
\frac{
U_{\mathrm{raw}}(x)-\langle U_{\mathrm{raw}}\rangle
}{
\operatorname{std}(U_{\mathrm{raw}})
}
\right],
\qquad
\alpha\sim\operatorname{Unif}[0.5,3].
\end{equation}
The resulting conservative field is computed as
\begin{equation}
\mathbf{F}(x)=-\boldsymbol{\nabla} U(x).
\end{equation}
Both force components are computed jointly and are not normalized or nonlinearly rescaled independently.

\end{document}